\documentclass[fleqn,usenatbib]{mnras}

\usepackage{newtxtext,newtxmath}

\usepackage[T1]{fontenc}

\DeclareRobustCommand{\VAN}[3]{#2}
\let\VANthebibliography\thebibliography
\def\thebibliography{\DeclareRobustCommand{\VAN}[3]{##3}\VANthebibliography}

\usepackage{graphicx}	
\usepackage{amsmath}	
\usepackage{booktabs}
\usepackage{cleveref}

\usepackage{totcount}
\newtotcounter{citnum} 
\def\oldbibitem{} \let\oldbibitem=\bibitem
\def\bibitem{\stepcounter{citnum}\oldbibitem}

\title{The atmospheric entry of cometary impactors}

\author[R. J. Anslow et al.]{Richard J. Anslow,$^{1}$\thanks{E-mail: \href{mailto:rja92@ast.cam.ac.uk}{rja92@ast.cam.ac.uk} (RJA)}
Amy Bonsor,$^{1}$
Zoe R. Todd,$^{2}$
Robin Wordsworth$^{3,4}$,
Auriol S. P. Rae$^{5}$,\newauthor
Catriona H. McDonald$^{1}$,
and Paul B. Rimmer$^{6}$
\\
$^{1}$Institute of Astronomy, University of Cambridge, Madingley Road, Cambridge, CB3 0HA, UK\\
$^{2}$Department of Chemistry, Department of Astronomy, University of Wisconsin-Madison, Madison, WI 53706, USA\\
$^{3}$School of Engineering and Applied Sciences, Harvard, Cambridge, MA 02138, USA\\
$^{4}$Department of Earth and Planetary Sciences, Harvard, Cambridge, MA 02138, USA\\
$^{5}$School of Geosciences, Grant Institute, University of Edinburgh, Edinburgh, EH9 3FE\\
$^{6}$Astrophysics Group, Cavendish Laboratory, University of Cambridge, JJ Thomson Ave, Cambridge, CB3 0HE, UK
}

\date{Accepted XXX. Received YYY; in original form ZZZ}

\pubyear{\the\year{}}

\begin{document}
\label{firstpage}
\pagerange{\pageref{firstpage}--\pageref{lastpage}}
\maketitle

\begin{abstract}
Cometary impacts play an important role in the early evolution of Earth, and other terrestrial exoplanets. Here, we present a numerical model for the interaction of weak, low-density cometary impactors with planetary atmospheres, which includes semi-analytical parameterisations for the ablation, deformation, and fragmentation of comets. Deformation is described by a pancake model, as is appropriate for weakly cohesive, low-density bodies, while fragmentation is driven by the growth of Rayleigh-Taylor instabilities. The model retains sufficient computational simplicity to investigate cometary impacts across a large parameter space, and permits simple description of the key physical processes controlling the interaction of comets with the atmosphere. We apply our model to two case studies. First, we consider the cometary delivery of prebiotic feedstock molecules. This requires the survival of comets during atmospheric entry, which is determined by three parameters: the comet's initial radius, bulk density, and  atmospheric surface density. There is a sharp transition between the survival and catastrophic fragmentation of comets at a radius of about 150\,m, which increases with increasing atmospheric surface density and decreasing cometary density. Second, we consider the deposition of mass and kinetic energy in planetary atmospheres during cometary impacts, which determines the strength and duration of any atmospheric response. We demonstrate that mass loss is dominated by fragmentation, not ablation. Small comets deposit their entire mass within a fraction of an atmospheric scale height, at an altitude determined by their initial radius. Large comets lose only a small fraction of their mass to ablation in the lower atmosphere.
\end{abstract}

\begin{keywords}
Meteors -- Comets -- Astrobiology
\end{keywords}



\section{Introduction}

The late delivery of volatile-rich, cometary impactors is often invoked as a mechanism responsible for key evolutionary processes that have shaped the composition of Solar system bodies. For example, cometary impacts have been suggested to play an important role in the evolution of Earth's atmosphere \citep{Chyba1990, Halliday2013, Marty2016}, in the atmospheric evolution of the outer Solar system satellites \citep{Zahnle1992_satellites, Sinclair2022}, and as a potential explanation for Jupiter's super-Solar metallicity \citep{Mahaffy2000, MullerHelled2024}. 

The evolution of Earth's atmosphere remains subject to continued debate, concerning the composition of late accreted material \citep{Halliday2013, Marty2016}, the physical mechanisms controlling atmospheric growth and erosion \citep{deNiem2012, Sinclair2020}, and potentially even the requirement for any volatile delivery at all \citep{Young2023}. 
Irrespective, however, of their role in delivering Earth's volatile inventory, cometary impacts have also been suggested to play a central role in establishing Earth's biosphere \citep{Ehrenfreund2002, Osinski2020}, and have been argued to play an important role in several prebiotic chemical scenarios \citep[e.g.,][]{Oro1961, Clark1988, Chyba1992, Sutherland2016}.

The cometary delivery of prebiotic molecules has re-emerged in recent years as a potentially viable, atmosphere-independent pathway towards the origin of life \citep{ToddOberg2020, Zellner2020}. This is, in part, motivated by the high concentrations of hydrogen cyanide (HCN) observed in Solar system comets \citep{MummaCharnley2011}. Not only this, but comets remain the only empirically-grounded source of cyanoacetylene in the concentrations required for prebiotic chemistry \citep{MummaCharnley2011}. Cyanoacetylene production in transiently reducing post-impact atmospheres remains several orders of magnitude below that used in prebiotic experiment \citep{Wogan2023}, whereas surface hydrothermal sources lack experimental support \citep{RimmerShorttle2024}.

Given HCN and cyanoacetylene are both key feedstock molecules in several prebiotic chemical scenarios \citep{Powner2009, Patel2015, Becker2019}, it is important to quantitively assess the ability of comets to deliver these molecules to local (subaerial) environments in high concentrations. Previous work has largely focussed on the ability of these (relatively fragile) molecules to survive the extremely high pressures and temperatures experienced by comets during hypervelocity impacts \citep{PierazzoChyba1999, ToddOberg2020}. This alone constrains successful cometary delivery to very low velocity impacts \citep[i.e., $\lesssim\,15\,{\rm km\,s}^{-1}$;][]{ToddOberg2020}. The ability of comets to survive atmospheric entry, and reach the surface intact, remains however relatively underexplored. This is also necessary for the concentration of these molecules in localised environments, and thus, the atmospheric break-up of comets has the potential to significantly reduce the number of environments able to support prebiotic chemistry \citep{Anslow2024}.

The potential importance of cometary impacts in the evolution of planetary environments is not limited to the Solar system; comets have been invoked as drivers of atmospheric evolution on rocky exoplanets \citep{Kral2018}, as sources of prebiotic feedstock molecules \citep{Anslow2023}, and as a pollutant of young hot Jupiters, potentially accounting for their enhanced metallicity \citep{Tsai2023, Zhang2023, SainsburyMartinezWalsh2024}. Recent work \citep{Sainsbury-Martinez2024} has also suggested that cometary impacts may generate observable features in the spectra of terrestrial exoplanets. This response is driven by the deposition of cometary water in the upper atmosphere, which provides a strong additional opacity source. Given that the location, and extent of mass loss from comets is dominated by deformation and fragmentation, not ablation \citep{Popova2019}, this atmospheric response is extremely sensitive to the cometary impact model.

The interaction of cometary impactors with planetary atmospheres is therefore a crucial physical process that must be accurately modelled in order to further constrain the cometary delivery of prebiotic feedstock molecules, and characterise the response of planetary atmospheres to cometary impacts.
Significant effort has been made characterising the atmospheric entry of meteoroids, which has diverged both in modelling approach (discussed in detail in \S\ref{sec:methods}), and physical application. These studies largely focus on asteroid breakup in Earth's atmosphere \citep[e.g.,][]{BaldwinSheaffer1971, Bronshten1983, HillsGoda1993, Svetsov1995}, and modelling crater field formation on Earth, Mars, and Venus \citep[e.g.,][]{PasseyMelosh1980, KorycanskyZahnle2005, Collins2022}. Significant progress has been afforded by the impact of comet Shoemaker-Levy 9 \citep{Sekanina1993}, the Tunguska impactor \citep{Chyba1993} and more recently the airburst over Chelyabinsk \citep{Popova2013}. Much of this progress was in the development of 2D, and 3D numerical models of meteoroid entry \citep{Ahrens1994, Boslough1994, ZahnleMacLow1994, Korycansky2000}, which are very computationally expense.

More recently, sophisticated hybrid models have been developed for asteroid breakup in order to characterise airburst altitudes, and the associated ground damage for risk assessments applications \citep{Register2017, Wheeler2017, Wheeler2018}. These models are successfully benchmarked against the Chelyabinsk meteor, but are restricted to asteroids in the size range 20-500\,m. 
The purpose of this work is to develop a numerical model appropriate for the atmospheric entry of weak, low density cometary impactors that is able to efficiently cover a wide range of parameter space\footnote{The \textsc{atmosentry} code is available to download at \url{https://github.com/richard17a/atmosentry}.}. This numerical model is validated against the progressive fragmentation models of \citet{Chyba1993} and \citet{HillsGoda1993}. 
To demonstrate the capabilities of the model, we focus on the following two applications:
\begin{enumerate}
    \item The survival of small comets, and delivery of prebiotic feedstock molecules.
    \item The deposition of mass and energy in the atmosphere during cometary impacts.
\end{enumerate}
The paper starts in \S\ref{sec:methods} by describing our numerical atmospheric entry model. In \S\ref{sec:results}, we present the outputs of the atmospheric entry model, and highlight the fundamental physical processes that control the interaction with, and therefore fate of comets in the atmosphere. In \S\ref{sec:discussion} we discuss the implications this has for both the cometary delivery of prebiotic feedstock molecules, and the atmospheric response to cometary impacts. Finally, \S\ref{sec:conclusion} summarises our conclusions.

\section{Model description}
\label{sec:methods}

There are three fundamental processes that dictate the interaction of comets with the atmosphere; deceleration (due to atmospheric drag), mass loss (due to ablation), and fragmentation. Here, we describe each process in turn.

\subsection{Trajectory through the atmosphere}
\label{sec:methods_traj}

We assume the meteoroid (or fragment thereof) arrives at the top of the atmosphere at an angle $\theta$ with respect to the local horizontal, velocity $\mathbf{v}=(v_x, v_y,v_z)$, radius $r$, and mass $m\left\{=4\pi\rho_m r^3/3\right\}$ (see figure~\ref{fig:model_schematic} for a schematic diagram illustrating the numerical model presented below). Atmospheric drag causes the deceleration of the meteoroid; assuming the standard prescription for drag at high Reynolds number, the meteoroid's velocity evolves as \citep[e.g.,][]{BaldwinSheaffer1971, PasseyMelosh1980, Bronshten1983},
\begin{equation}
    \label{eq:drag_eqn}
    m\dfrac{d\mathbf{v}}{dt} = -\dfrac{1}{2}C_D\rho_{\rm atm}(z)A|\mathbf{v}|\mathbf{v} -g(z)\mathbf{\hat{e}}_z,
\end{equation}
where $C_D$ is the meteoroid's (dimensionless) drag coefficient \citep[$\sim\,$$0.5$; e.g.,][]{PasseyMelosh1980}, and $A$ its cross-sectional area. Both the atmospheric density, $\rho_{\rm atm}(z)$, and gravitational acceleration, 
\begin{equation}
    g(z) = \frac{GM_{\rm pl}}{(R_{\rm pl} + z)^2},
\end{equation}
are functions of altitude, $z$, and $M_{\rm pl}$ ($R_{\rm pl}$) is the mass (radius) of the planet. The meteoroid's position, $\mathbf{x}=(x,y,z)$, is recorded throughout the simulation (allowing for the calculation of impact location on the planet's surface), which evolves simply as
\begin{equation}
    \dfrac{d\mathbf{x}}{dt} = \mathbf{v}.
\end{equation}

\begin{figure*}
    \centering
    \includegraphics[width=1\textwidth]{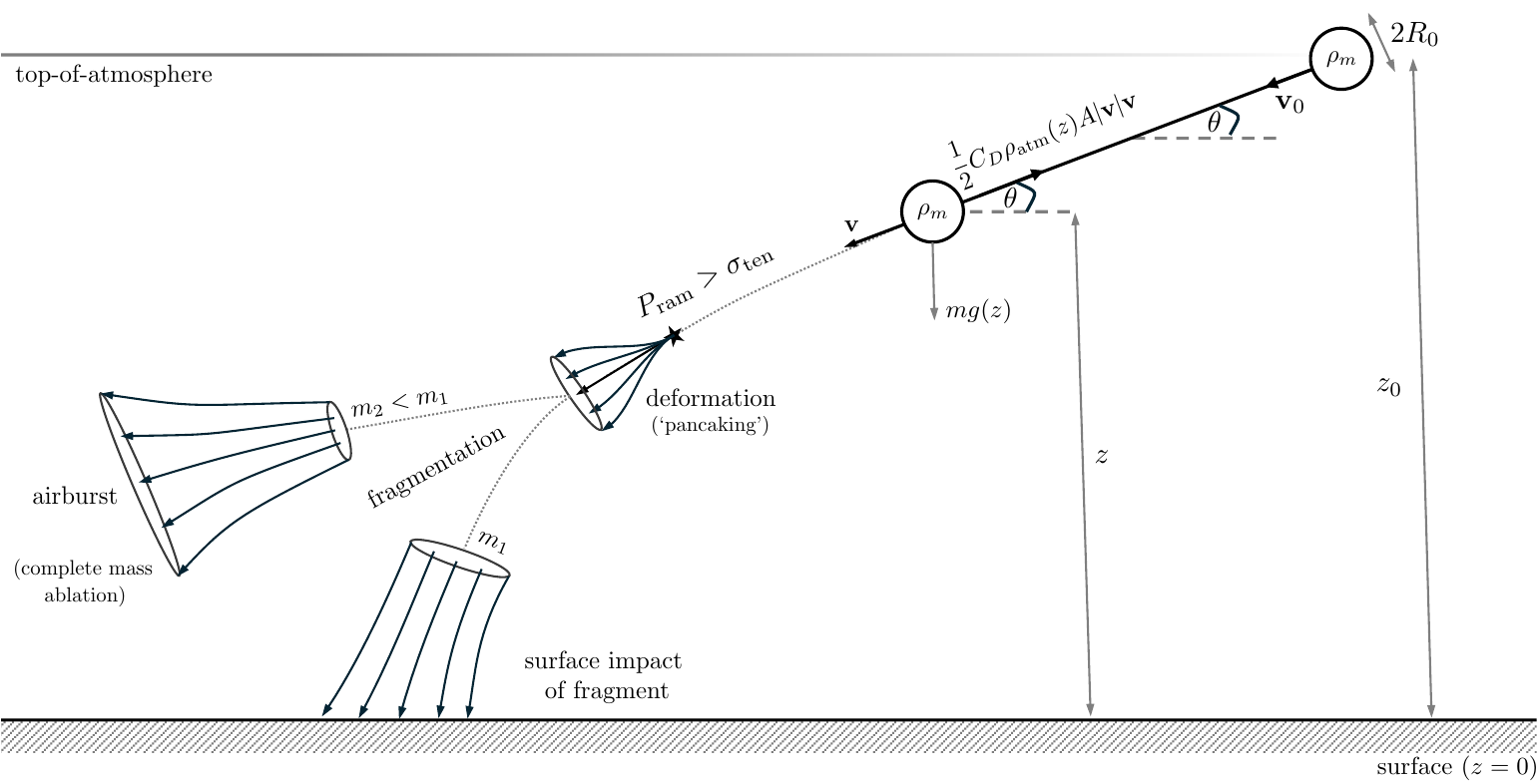}
    \caption{Schematic diagram describing the numerical model presented in \S\ref{sec:methods}. The comet arrives at the top of the atmosphere at an altitude $z_0$, with initial radius $R_0$ and velocity $\mathbf{v}_0$. The comet begins to deform when the ram pressure exceeds its tensile strength, which is indicated by the star. The radius of the comet increases quickly after this point (expanding laterally into a cylindrical shape), causing its rapid deceleration. After the growth of the Rayleigh-Taylor (hydrodynamic) instability, it fragments into two child fragments with masses $m_2 < m_1 < M_0$. These continue to deform quickly through the lower atmosphere, with the smaller fragment (mass $m_2$) surrendering the entirety of its mass to the surrounding atmosphere via ablation. The larger fragment (mass $m_1$) is able to reach the surface (heavily deformed, and at a speed much less than $|\mathbf{v}_0|$).}
    \label{fig:model_schematic}
\end{figure*}

\subsection{Mass ablation}
\label{sec:methods_mass_ablation}

During atmospheric entry, a very high temperature shock front forms at the leading edge of the meteoroid. The radiation from this atmospheric shock front drives the progressive mass ablation of the meteoroid. The rate of mass loss, following the classical ablation equation of \citet{Bronshten1983}, is given by
\begin{equation}
    \xi\dfrac{dm}{dt} = -\dfrac{1}{2}C_H\rho_{\rm atm}Av^3,
\end{equation}
where $\xi$ is the heat of ablation (the energy required to heat the surface to the required temperature for ablation), and $C_H$ the (dimensionless) heat transfer coefficient, which describes the partition of energy transfer between the comet and surrounding atmosphere.

Observations indicate that $C_H$ remains roughly constant in the upper atmosphere, but varies inversely with atmospheric density below $\sim\,$30\,km \citep{Biberman1980}. The ablation of larger meteoroids in the lower atmosphere is attributed to the absorption of the radiative flux from the shock front, with the temperature of the shocked gas strongly regulated by thermal ionization. We therefore introduce an upper limit on the ablation rate \citep[following][]{Zahnle1992} given by, 
\begin{equation}
    \label{eq:mass_ablation}
    \xi\dfrac{dm}{dt} = -A\,\text{min}\left(\dfrac{1}{2}C_H\rho_{\rm atm}v^3,\sigma_{\rm SB}T^4\right),
\end{equation}
where $T\sim25,000\,$K is the temperature of the shocked gas, and $\sigma_{\rm SB}$ the Stefan-Boltzmann constant. 

It is clear from equation~\ref{eq:mass_ablation} that the rate of ablation (particularly in the upper atmosphere) is strongly dependent on the heat transfer coefficient; this coefficient is, however, poorly constrained, and is sensitive to the velocity of the meteoroid, its shape, and the flow regime. Neglecting further complications, which include the shielding effects of vapor, and re-radiation from the meteoroid's leading edge, \citet{Svetsov1995} demonstrated that $C_H$ changes by orders of magnitude ($\sim\,$$10^{-5}-10^{-1}$) depending only on the meteoroid's size. This uncertainty therefore has the capacity to bias estimates of mass loss from large comets by several orders of magnitude. 

The impact of comet Shoemaker-Levy 9 into Jupiter's atmosphere therefore provides an important observational constraint for km-scale bodies, and supports a lower value of $C_H\sim\,$$10^{-3}$. Larger values \citep[$\sim\,$0.1, valid only for cm-sized bodies;][]{Bronshten1983} significantly overestimate mass ablation, and correspondingly underestimate the comet's penetration depth into the Jovian atmosphere \citep[e.g.,][]{Ahrens1994, Svetsov1995}. This has been subsequently supported by high-resolution simulations, which include both coupled radiation and ablation, and predict values significantly less than 0.1 \citep{Johnston2018}. We expect that $C_H$ will also change with atmospheric composition, given this will strongly influence the radiative transfer of energy. However, to the best of our knowledge, we are not aware of any study that quantitatively investigates this dependence.
We therefore set $C_H=10^{-3}$ throughout \citep[see also,][]{ZahnleMacLow1994, FieldFerrara1995}.

\subsection{Deformation and fragmentation}
\label{sec:deformation_fragmentation}

During atmospheric entry, the meteoroid will experience an exponentially increasing ram pressure at its leading edge, as a consequence of increasing atmospheric density. This will, at some point, exceed its material strength. Not all bodies will however be affected equally \citep[e.g.,][]{Melosh1989}; encountering a column-integrated atmospheric mass $P_{\rm surf}/g\sin{\theta}$, we expect that (by conservation of momentum) only comets smaller than 
\begin{equation}
    \label{eq:qual_rcrit}
    r_{\rm crit} \approx 100~\left(\frac{P_{\rm surf}}{1\,{\rm bar}}\right) \left(\frac{\rho_m}{0.4\,{\rm g\,cm}^{-3}}\right)^{-1}\left(\frac{g}{9.81\,{\rm m\,s}^{-2}}\right)^{-1} \left(\frac{\sin{\theta}}{1/\sqrt{2}}\right)^{-1}\,{\rm m}
\end{equation}
will significantly deform, and fragment\footnote{A similar argument is made in \citet{Svetsov1995}, who compare the sound-crossing time, $2r/c$, where $c$ is the speed of sound in the body, with the characteristic dynamical time through the atmosphere, $H/v\sin{\theta}$. Assuming $c\approx400\,{\rm m\,s}^{-1}$ \citep[accounting for the inverse correlation of sound speed with porosity;][]{Flynn2018}, $v=20\,{\rm km\,s}^{-1}$, and $H=7\,$km, this predicts $r_{\rm crit}\sim100\,m$, in close agreement with equation~\ref{eq:qual_rcrit}.}. The deformation of small bodies spreads their mass over a larger area, thus increasing atmospheric drag, and in turn driving enhanced deformation. This feedback leads to the rapid deposition of the meteoroid's kinetic energy within a fraction of an atmospheric scale height, in an explosive `airburst'.

The onset of fragmentation is determined by the meteoroid's altitude (atmospheric density), velocity and strength. The appropriate material strength to adopt is, however, subject to debate given that several processes are potentially responsible for the onset of fragmentation \citep{Svetsov1995}. Fragmentation may occur either when the stagnation pressure of the atmospheric flow exceeds the material's compressive strength,
\begin{equation}
    \rho_{\rm atm} v^2 > \sigma_{\rm com},
\end{equation}
or when the average ram pressure exceeds its tensile strength,
\begin{equation}
    \label{eq:frag_tensile_strength}
    \frac{C_D}{2}\rho_{\rm atm} v^2 > \sigma_{\rm ten}.
\end{equation}
Comets are extremely weak, and estimates of their compressive and tensile strength (approximately) coincide within error \citep{Groussin2019}. The onset of fragmentation will therefore, unavoidably, occur at high altitude (except in the most tenuous atmospheres) regardless of which break-up condition is used. Here, we determine the onset of fragmentation using equation~\ref{eq:frag_tensile_strength}, and assume a characteristic tensile strength $\sigma_{\rm ten}\approx10^4\,$Pa \citep{Groussin2019}\footnote{Note, this is an intermediate value between that of pure ice \citep[$\sim\,$$10^6\,$Pa;][]{Petrovic2003}, which likely overestimates the true strength of porous cometary nuclei, and observational estimates ($\sim\,$$10^3\,$Pa) derived from the tidal disruption of comet Shoemaker-Levy 9 \citep{BenzAsphaug1994}.}. We discuss the sensitivity of our results to this choice in \S\ref{sec:results_comet_fates}.

\subsubsection{Deformation}

There exist a variety of analytical methods that can be used to model a meteoroid's subsequent fragmentation, which are split into two main approaches \citep{ArtemievaShuvalov2001}. Pancake models \citep{HillsGoda1993, Zahnle1992, Chyba1993}, which assume the parent body is heavily fragmented, modelling the continuous deformation of an incompressible fluid, and discrete fragmentation models \citep{BaldwinSheaffer1971, PasseyMelosh1980}, which model the successive fragmentation of the parent body into individual pieces. Given the very low material strength, and bulk density of cometary nuclei \citep{Groussin2019}, pancake models most faithfully describe their deformation.

Pancake models rely on the large imbalance between the ram pressure experienced by the meteoroid's leading edge, and the atmospheric pressure against the side walls. In response to the induced pressure gradients, a transverse force drives the lateral deformation of the meteoroid into a cylinder of height $h=m/\pi \rho_m r^2$ (implicitly assuming that its density, $\rho_m$, remains constant). Its radius is assumed to increase according to \citep{Zahnle1992, Chyba1993},
\begin{equation}
    \label{eq:deformation_d2r}
    r\dfrac{d^2r}{dt^2} = \dfrac{C_D}{2}\left(\dfrac{\rho_{\rm atm}}{\rho_{\rm m}}\right)v^2,
\end{equation}
where $r$ is the meteoroid's radius, and $\rho_m$ its bulk density.

This deformation will not, however, continue indefinitely, with the implicit assumption of a collective bow-shock eventually breaking down. Several numerical studies have demonstrated this is driven by the growth of hydrodynamic instabilities, of which the Rayleigh-Taylor instability appears most salient \citep[e.g.,][]{FieldFerrara1995, Crawford1997, Korycansky2000, Korycansky2002}. 

Following \citet{KorycanskyZahnle2005}, the deformation of the meteoroid is stopped after two Rayleigh-Taylor growth timescales ($N_{\rm RT}$). The growth rate of a perturbation of wavenumber $k$ is given by \citep{FieldFerrara1995},
\begin{equation}
    \label{eq:N_RT}
    \dfrac{dN_{\rm RT}}{dt} = \left(\dfrac{3}{8}C_D\dfrac{\rho_{\rm atm}(z)}{\rho_m}\dfrac{k}{r}\right)^{1/2}v(z),
\end{equation}
and in this study we consider only the most destructive mode, which is assumed to have wavenumber $k\sim\pi/r$ \citep{Korycansky2000, KorycanskyZahnle2005}. At this point, after the growth of the Rayleigh-Taylor instability, high-resolution numerical simulations indicate that distinct fragments then develop their own bow shocks, and move apart \citep[e.g.,][]{Korycansky2000, Korycansky2002}. This is consistent with both the morphology of crater fields \citep[e.g.,][]{PasseyMelosh1980}, and direct observation of bolides entering Earth's atmosphere \citep{BorovickaSpurny1996}.
We describe the properties of these child fragments next.
\begin{figure*}
    \centering
    \includegraphics[width=0.85\textwidth]{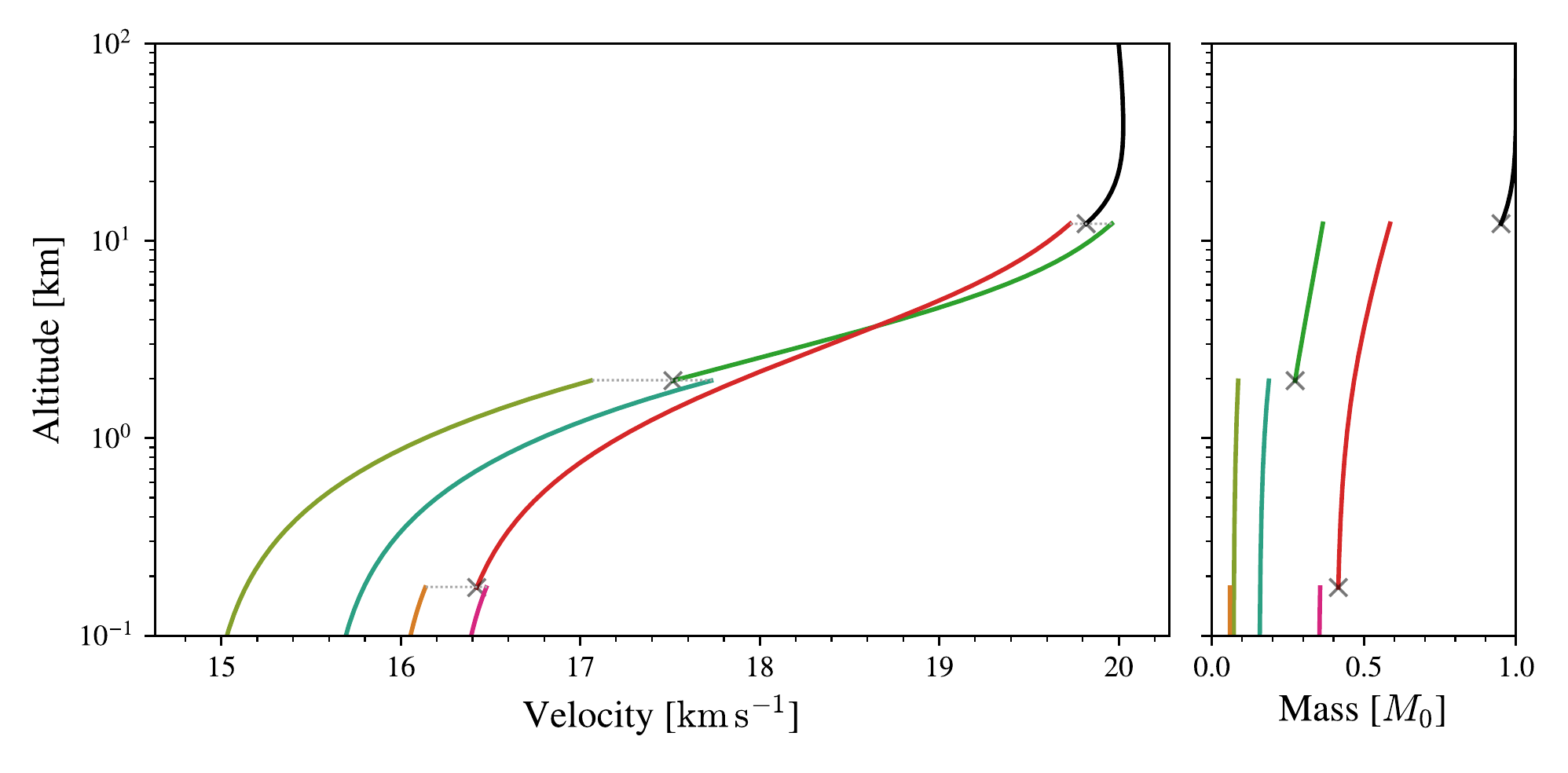}
    \caption{An example trajectory of a 150\,m comet (with initial velocity $20\,{\rm km\,s}^{-1}$), and its child fragments, through Earth's atmosphere. Black crosses indicate the point of fragmentation, after which child fragments are generated according to the prescription described in \S\ref{sec:fragment_masses}-\ref{sec:fragment_vels}. Colors are used to highlight the connection between parent, and child fragments. \textbf{Left panel:} The comet's deceleration, as a function of altitude. \textbf{Right panel:} The comet's mass loss, as a function of altitude. The stochasticity in child fragment masses, and corresponding asymmetry in separation velocities is seen in both the inset panel (\textbf{left}), and mass evolution panel (\textbf{right}).}
    \label{fig:zoom_in_panel}
\end{figure*}

\subsubsection{Fragment masses}
\label{sec:fragment_masses}

The number of fragments ($n$) produced during fragmentation remains poorly constrained, however numerical simulations \citep{Korycansky2002}, and groups of craters on Venus \citep{HerrickPhillips1994} suggest this number is in the range $2-4$. The masses of the child fragments are chosen proportional to a random variable,
\begin{equation}
    \frac{m_i}{m_{\rm parent}} = x, \hspace{2em} \text{where} ~ x\sim\mathcal{U}[0,1]
\end{equation} 
and normalised such that
\begin{equation}
    \sum_{i=1}^n{m_i} = m_{\rm parent},
\end{equation}
where $m_{\rm parent}$ is the mass of the meteoroid at the point of fragmentation, and $m_i$ the masses of the $n$ child fragments. For simplicity, in this study we assume only 2 child fragments are produced \citep[following e.g.,][]{Collins2022}, but we demonstrate the robustness of our results to this assumption in appendix~\ref{appendix:tables}.

\subsubsection{Fragment separation velocities}
\label{sec:fragment_vels}

As identified in \citet{PasseyMelosh1980}, during the onset of fragmentation the interaction of newly-formed bow shocks supplies a transversal velocity component to the child fragments. The characteristic speed of their separation is given by\footnote{Their analysis assumes a constant acceleration ($\alpha$) until the fragments reach a separation of order $r$. Their final transverse velocity is thus $v_t\sim (\alpha r)^{1/2}$. This acceleration is driven by the large dynamic pressure at the leading edge ($\rho_{\rm atm} v^2$), and so $\alpha \sim (\rho_{\rm atm}/\rho_m)(v^2/r)$. Neglecting constants (order unity), the transverse velocity is therefore $v_t \simeq (\rho_{\rm atm}/\rho_m)^{1/2}v(z)$. The model parameter $C_V$ is constrained via observations of crater fields on Venus \citep{KorycanskyZahnle2005}, and Mars \citep{Collins2022}.}
\begin{equation}
    \label{eq:separation_vel}
    v_s = \left(C_V\dfrac{\rho_{\rm atm}}{\rho_m}\right)^{1/2}v(z),
\end{equation}
where $v(z)$ is the parent fragment’s speed at the point of fragmentation, and $C_V$ a numerically determined constant of proportionality \citep{PasseyMelosh1980, ArtemievaShuvalov2001}. This constant will only affect the distribution of field craters; we assume $C_V\approx0.5$ \citep{Collins2022}.

Following \citet{KorycanskyZahnle2005}, the transverse velocity $\mathbf{v}_{t,i}=(v_{\alpha i}, v_{\beta i})$ supplied to individual child fragments is determined by a random azimuth, $\phi_i$. To ensure we conserve total linear momentum, the transverse velocity is given by
\begin{equation}
    \mathbf{v}_{t,i} = v_s\left(\hat{\mathbf{n}}_i - \dfrac{1}{m_{\rm parent}} \sum_{j=1}^n{m_j\mathbf{\hat{n}}}_j\right),
\end{equation}
where $\hat{\mathbf{n}}_i=(\sin{\phi_i}, \cos{\phi_i})$. The initial velocity, $\mathbf{v}_i$, of each child fragment $i$ is thus,
\begin{equation}
    \mathbf{v}_i = \mathbf{v}_{\rm parent} + 
    \begin{pmatrix}
        v_{\beta i} \cos{\theta} \\
        v_{\alpha i} \\
        v_{\beta i} \sin{\theta}
    \end{pmatrix},
\end{equation}
where $\mathbf{v}_{\rm parent}$ is the parent fragment's velocity at the point of break-up, and $\theta$ its angle to the local horizontal. 

The child fragments' trajectories are then calculated (also subject to drag, mass ablation, and deformation) until either they reach the surface, or they themselves fragment into $n$ further child fragments.
\begin{figure*}
    \centering
    \includegraphics[width=1\textwidth]{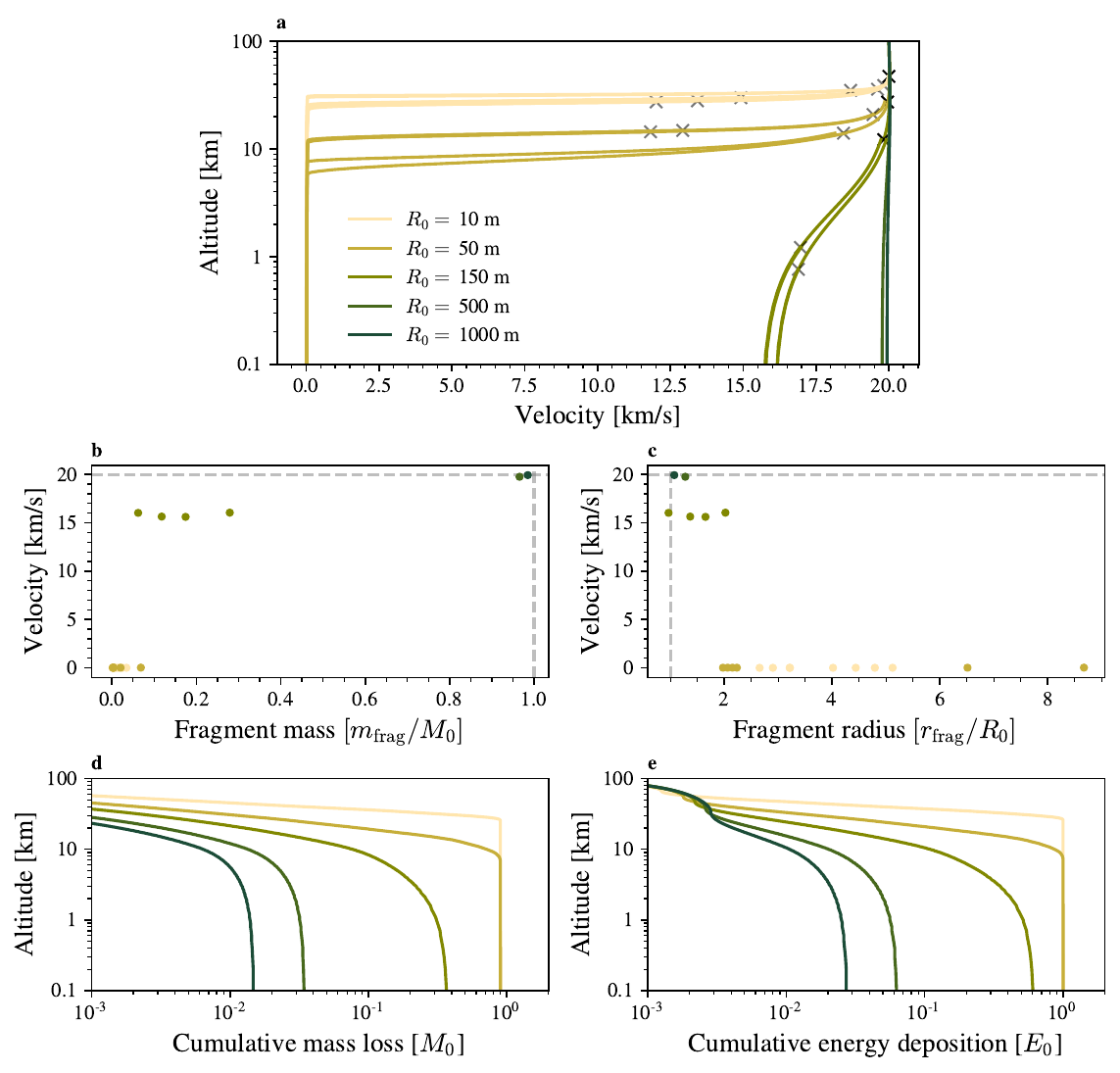}
    \caption{Key parameters describing the effects of atmospheric entry are shown for comets with initial radii in the range $10-1000\,$m, which is the key parameter determining the relative importance of atmospheric entry. \textbf{(a)} Effective atmospheric deceleration is only possible for the smallest comets (i.e., $R_0\lesssim150\,$m), in agreement with the qualitative argument presented in \S\ref{sec:deformation_fragmentation}. \textbf{(b)} The mass and velocity of surviving fragments that impact Earth's surface. Surviving fragments from $R_0 \lesssim\,150\,$m comets carry only a very small fraction of the comet's initial mass, and impact the surface at (roughly) terminal velocity. Larger comets reach the surface without significant deceleration, or mass loss. \textbf{(c)} Large comets reach the surface without significant deformation, in stark contrast with smaller bodes that experienced significant lateral expansion. \textbf{(d)} Significant mass loss (approaching 100\%) is observed for small comets, occurring at altitudes proportional to initial size. \textbf{(e)} Energy deposition in the atmosphere, due to both deceleration, and mass loss (normalised by the comet's initial kinetic energy). The initial mass, radius, and velocity of the comet are highlighted via the gray dashed lines in panels \textbf{c}, and \textbf{d}.}
    \label{fig:comet_trajectory_gallery}
\end{figure*}
Note, the rate of successive fragmentation accelerates rapidly, given that their mass is already spread over a very large cross-sectional area. As discussed, this increases the strength of aerodynamic drag, mass ablation, and the rate of further deformation.

\begin{figure*}
    \centering
    \includegraphics[width=\linewidth]{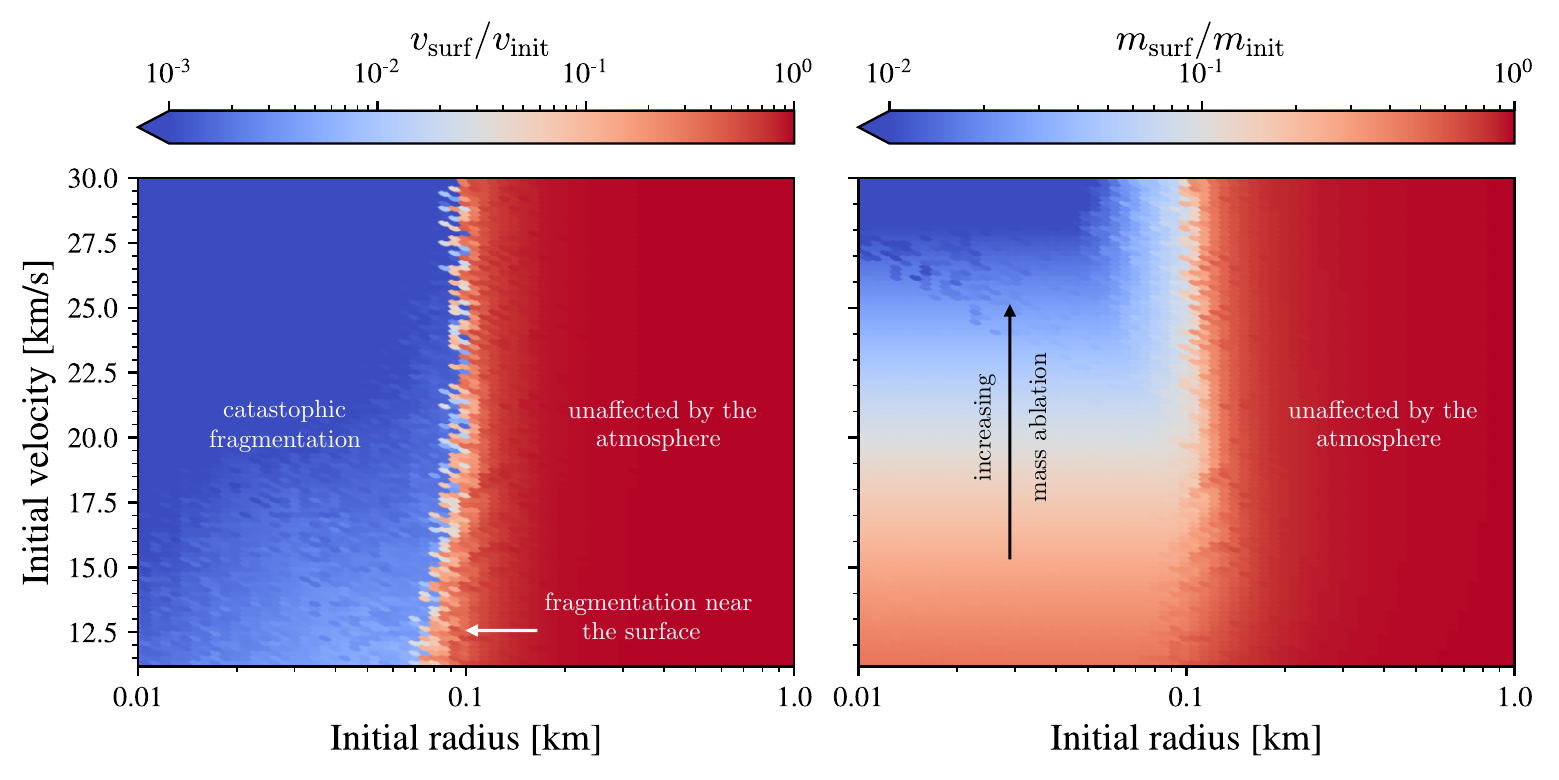}
    \caption{
    The effects of atmospheric entry on comets' impact velocity at the surface (\textbf{left}), and remaining mass (\textbf{right}) is shown as a function of initial radius, and velocity for an Earth-like atmosphere (with surface density $1.225\,{\rm kg\,m}^{-3}$). For comets that fragment (i.e., with small initial radius), this is given as the (mass-weighted) average velocity, and total mass of fragments that reach the surface. There is a sharp transition in the response to the atmosphere at an initial radius of about 100\,m, in agreement with analytical predictions (equation~\ref{eq:qual_rcrit}). 
    }
    \label{fig:heatmaps}
\end{figure*}

\section{Results}
\label{sec:results}

\subsection{An example comet's trajectory}
\label{sec:results_example_traj}

As described in \S\ref{sec:methods}, the interaction of a comet with the atmosphere causes its deceleration, driven by the transfer of momentum from the incoming atmospheric flow, and mass-loss, driven by the transfer of kinetic energy from the atmospheric flow. Both atmospheric drag and mass ablation are proportional to the surrounding atmospheric density (see \cref{eq:drag_eqn,eq:mass_ablation}), and are therefore relatively inefficient in the tenuous upper atmosphere. This is seen in figure~\ref{fig:zoom_in_panel}, which shows the trajectory of a 150\,m comet through Earth's atmosphere; the mass, and velocity of the comet remain roughly constant throughout the first 50\,km of the atmosphere.

Whilst the velocity of the comet remains roughly constant during this initial descent, the ram pressure ($P_{\rm ram}=\rho_{\rm atm}v^2$) at its leading edge rapidly increases, as a consequence of the exponentially increasing atmospheric density. When the ram pressure exceeds the material strength of the comet \citep[of the order $10^4\,$Pa;][]{Groussin2019}, it will begin to deform, and fragment. This will occur high ($>50\,$km) in Earth's atmosphere for comets of all sizes. The motion of the comet after this point is characterised by significant deceleration and mass-loss.

The deforming comet begins to distribute its mass over a larger cross-sectional area, increasing both the atmospheric drag and rate of mass ablation, thereby depositing its energy into the surrounding atmosphere at a rapidly increasing rate. This is seen in figure~\ref{fig:zoom_in_panel} between $50-10\,$km, where we see, for the first time, the significant deceleration, and ablation of the comet. This deformation continues until the onset of fragmentation, which is controlled by the growth of the (hydrodynamical) Rayleigh-Taylor instability, with timescale\footnote{Adopting an average atmospheric density of about $0.05\,{\rm kg\,m}^{-3}$ between $50-10\,$km, appropriate for an Earth-like atmosphere with scale height $7\,$km.} (see equation~\ref{eq:N_RT}),
\begin{align}
    \label{eq:RT_growth_timescale}
    \tau_{\rm RT} \sim 1 &\left(\dfrac{\rho_m}{0.6\,{\rm g\,cm}^{-3}}\right)^{1/2}  \left(\dfrac{\rho_{\rm atm}}{0.05\,{\rm kg\,m}^{-3}}\right)^{-1/2} \times \nonumber\\ 
    &\left(\dfrac{v}{20\,{\rm km\,s}^{-1}}\right)^{-1} \left(\dfrac{r}{150\,{\rm m}}\right)\,{\rm s}.
\end{align}
Over the course of two Rayleigh-Taylor growth timescales, the comet descends a total distance $\Delta z \sim 2 \, \tau_{\rm RT} v\, \mathrm{cosec}{\theta} \sim 30\,$km, in rough agreement with figure~\ref{fig:zoom_in_panel}.

After the growth of the destructive Rayleigh-Taylor instability, at an altitude of about 10\,km, the comet splits into two child fragments (green and red curves, figure~\ref{fig:zoom_in_panel}). These differ in both initial mass (see \S\ref{sec:fragment_masses}), and transverse velocity (see \S\ref{sec:fragment_vels}), and immediately continue deforming in response to the extreme ram pressure experienced at their leading edge. Both fragments decelerate substantially during this final 10\,km of descent through the atmosphere, surrendering the majority of the comet's initial kinetic energy to the surrounding atmosphere. 

The rate of deformation, and growth of hydrodynamical instabilities, in the dense lower atmosphere is much quicker for the smaller (green) child fragment (see equation~\ref{eq:RT_growth_timescale}). The smaller fragment therefore fragments first, in this case about 1\,km above the surface. The fragments have little time to move apart downrange following initial break-up at 10\,km, and so the impact craters formed once they reach the surface will likely overlap, forming a crater field. The diameter of this crater field is dominated by the largest fragment (pink, figure~\ref{fig:zoom_in_panel}), which reaches impacts surface at maximum velocity.

\subsection{What controls the fate of comets in the atmosphere?}
\label{sec:results_comet_fates}

As described in \S\ref{sec:results_example_traj}, during atmospheric entry a comet begins to deform when the ram pressure at its leading edge exceeds its material strength. In an isothermal atmosphere (with constant scale height), this will occur at a characteristic altitude (see equation~\ref{eq:frag_tensile_strength})
\begin{equation}
    \label{eq:break_up_altitude}
    z_{\rm break} \simeq \log{\left(\dfrac{\rho_{\rm atm, 0} v_0^2}{\sigma_{\rm ten}}\right)} H,
\end{equation}
where $\rho_{\rm atm, 0}$ is the atmospheric surface density, $H$ the atmospheric scale height, $v_0$ the comet's initial velocity, and $\sigma_{\rm ten}$ the comet's tensile strength. Whilst the tensile strength of cometary nuclei is poorly constrained, encompassing both 1\,Pa \citep[as inferred from overhangs on comet 67P;][]{Attree2018}, and $10^6\,$Pa \citep[the tensile strength of pure water ice;][]{Petrovic2003}, this break-up height only varies in the range of $20-60\,$km. In other words, comets will always begin to break up at high altitudes in Earth-like atmospheres. The exact value of the tensile strength, within this wide range, has little effect on the survival of comets.

The fate of comets is instead determined by the subsequent rate of deformation, and the growth of destructive hydrodynamical instabilities. When the growth timescale of the Rayleigh-Taylor instability (equation~\ref{eq:RT_growth_timescale}) exceeds the dynamical atmosphere-crossing timescale ($\sim H/v\sin{\theta}$), the comet will reach the surface before fragmentation. The survival of comets is therefore controlled by the following three factors:
\begin{enumerate}
    \item the comet's impact parameters, specifically radius, velocity, and angle of incidence at the top of the atmosphere,
    \item the comet's bulk density,  
    \item the planet's atmospheric surface density.
\end{enumerate}
We demonstrate the importance of these factors for the fate of comets during atmospheric entry in the remainder of this Section.

\subsubsection{The comet's impact parameters}
\label{sec:discussion_comet_impact_params}

During atmospheric entry, the fate of a comet is primarily determined by its initial radius, as is demonstrated in figure~\ref{fig:comet_trajectory_gallery}~(a-c). Small comets, with initial radius less than 50\,m, encounter an atmospheric column more massive than itself and deform rapidly, losing both their mass and kinetic energy (at high altitude) to the surrounding atmosphere. In contrast, comets with initial radius larger than $\sim\,$$500\,$m reach the surface without the significant loss of mass, or kinetic energy to the atmosphere, and reach the surface with their initial velocity essentially unchanged. 

In addition to controlling the survival of comets, their initial size also determines the location, and extent of mass loss in the atmosphere.
It is clear from figure~\ref{fig:comet_trajectory_gallery}~(d-e) that mass (and energy) loss is dominated by fragmentation, with the rapid increase in mass ablation during deformation orders of magnitude larger than in the upper atmosphere. Small comets therefore deposit essentially all of their initial mass, and energy into the surrounding atmosphere within a fraction of an atmospheric scale height. Larger comets, on the other hand, lose only a small fraction of their initial mass to ablation, which occurs primarily in the dense lower atmosphere. We note that negligible mass loss due to ablation is independently supported by observations of meteoroids in Earth's atmosphere \citep{Popova2019}.

The transition between the survival, and catastrophic fragmentation of comets in the atmosphere is rapid, and independent of initial velocity (see figure~\ref{fig:heatmaps}). This transition occurs at approximately $150\,$m in a 1\,bar (i.e., Earth-like atmosphere), which is in good agreement with qualitative expectations based on the conservation of momentum (equation~\ref{eq:qual_rcrit}). The effects of initial velocity are only apparent in the total mass of fragments reaching the surface, with the $v^3$ dependence of mass ablation causing the near-total evaporation of small, high velocity comets. This is a consequence of the increased transfer of kinetic energy from the atmospheric flow.

\begin{figure}
    \centering
    \includegraphics[width=\linewidth]{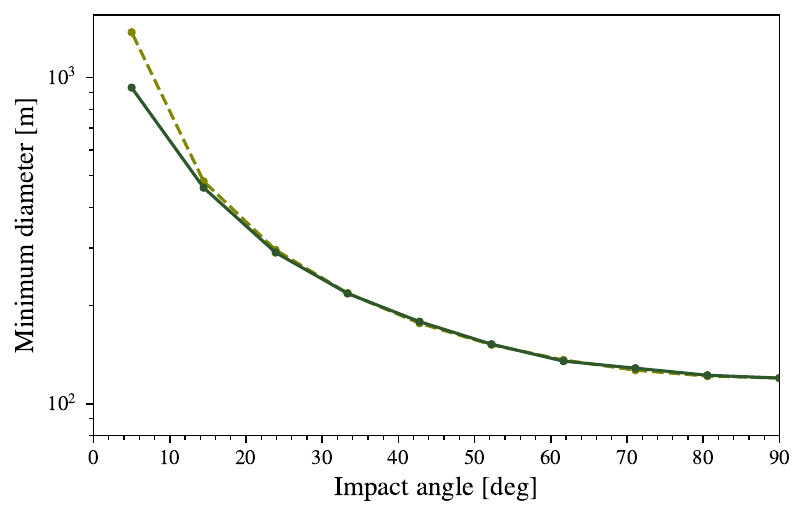}
    \caption{The minimum cometary diameter required to survive atmospheric entry as a function of entry angle (assuming Earth-like atmospheric surface density). The dashed line corresponds to the $\mathrm{cosec}{\theta}$ scaling, in agreement with simple analytical expectation (see \S\ref{sec:discussion_comet_impact_params}). Comets are assumed to survive atmospheric entry if the total mass of fragments reaching the surface exceeds 75\% of the comet's initial mass.}
    \label{fig:minimum_cometary_diameter_angles}
\end{figure}

Whilst the initial velocity of a comet has little effect on its interaction with the atmosphere, the angle of incidence at the top of the atmosphere makes a significant difference to its survival (figure~\ref{fig:minimum_cometary_diameter_angles}). This is caused by the increased path length through the atmosphere, which drives the catastrophic fragmentation of (significantly) larger comets. The sharp transition seen in figure~\ref{fig:heatmaps}, separating the survival and disruption of comets in the atmosphere, therefore changes with angle, scaling with $\mathrm{cosec}\,{\theta}$ (the increase in atmospheric path length). Given that comets arrive at the top of a planet's atmosphere with a range of angles, following the differential probability distribution \citep{Shoemaker1962}
\begin{equation}
    dP = \sin{2\theta}d\theta,
\end{equation}
the radius of the smallest comets able to survive atmospheric entry will also vary. This minimum size increases sharply from $\sim\,$$150\,$m (corresponding to the most probable angle, $45^\circ$) to almost 1\,km, for angles below $20^\circ$. We note, however, that such impacts will be relatively uncommon, accounting for less than roughly $10\,$\% of total impacts. Given the much wider range in the size of cometary nuclei in comparison to possible angles of incidence, the dominant effect controlling the survival of comets during atmospheric entry is initial radius.

\subsubsection{The density of cometary nuclei}
\label{sec:discussion_comet_density}

\begin{figure}
    \centering
    \includegraphics[width=\linewidth]{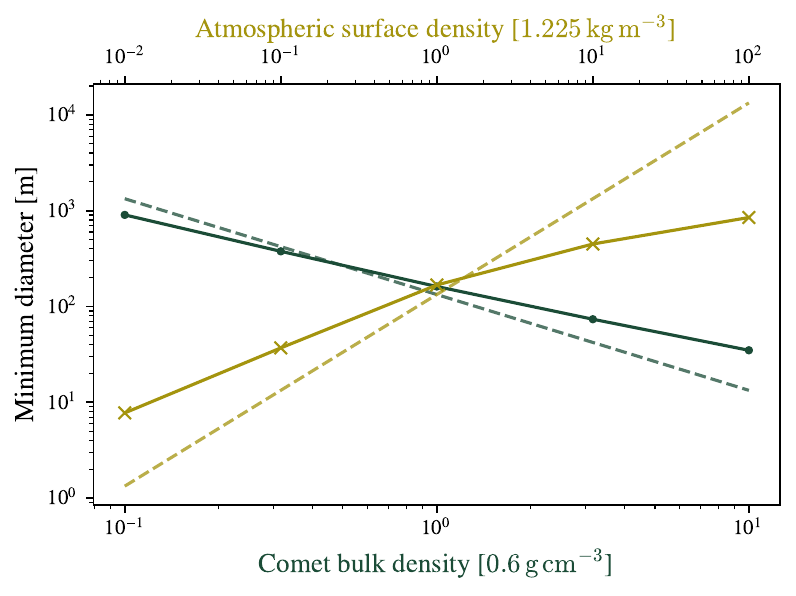}
    \caption{The minimum cometary diameter required to survive atmospheric entry as a function of the comet's bulk density (circles), and atmospheric surface density (crosses). Analytical estimates of this minimum size, using equation~\ref{eq:qual_rcrit}, are shown above as dashed lines. This qualitative estimate is able to account for the dependence on comet bulk density to surprising quantitative accuracy. These predictions are, however, less accurate for extreme values of atmospheric surface density, which is due to the fact that changes in surface density not only affect the rate of deformation and fragmentation, but also the altitude at which deformation begins (see equation~\ref{eq:break_up_altitude}). Comets are assumed to survive atmospheric entry if the total mass of fragments reaching the surface exceeds 75\% of the comet's initial mass.}
    \label{fig:minimum_cometary_diameter}
\end{figure}

The density of cometary nuclei are inferred to vary substantially \citep[with many comets consistent with bulk densities much lower than $0.6\,{\rm g\,cm}^{-3}$;][]{Kokotanekova2017}, which strongly controls the fate of comets in the atmosphere for two primary reasons. First, low density comets are less massive than their high density counterparts, and will therefore be more easily stopped by the atmosphere (simply due to momentum conservation). Second, the rate of deformation, and growth of the Rayleigh-Taylor instability are both greater for low density comets. This, in turn, increases the strength of aerobraking and mass ablation, causing the break up of even relatively massive low density comets. This is therefore a twofold effect\footnote{We note also that low density (i.e., highly porous) comets will typically have lower material strength than higher density comets, which will be made of more coherent blocks of rock and ice \citep[e.g.,][]{Collins2005}. Given however the relatively small effect tensile strength has on cometary survival (see \S\ref{sec:results_comet_fates}), we expect this to be a much smaller effect.}, such that low density comets must be much larger to avoid catastrophic fragmentation at high altitude. This trend is seen clearly in figure~\ref{fig:minimum_cometary_diameter}, which demonstrates the sharp increase in minimum cometary diameter with decreasing bulk density. Only comets larger than $\sim\,$1\,km will survive atmospheric entry for bulk densities below $\sim\,$$0.1\,{\rm g\,cm}^{-3}$.

\subsubsection{The atmospheric surface density}

Finally, the atmospheric surface density is itself an key parameter, with dense planetary atmospheres able to successfully decelerate, and catastrophically fragment much larger (i.e., km-scale) comets (figure~\ref{fig:minimum_cometary_diameter}). This can be understood relatively straightforwardly, given that increases in atmospheric surface density are directly proportional to increases in the column-integrated atmospheric mass a comet will encounter during its descent (see equation~\ref{eq:qual_rcrit}). A rapid increase in minimum cometary diameter required to survive atmospheric entry as a function of surface density can be seen clearly in figure~\ref{fig:minimum_cometary_diameter}.

The interaction of small comets with dense atmospheres are particularly dramatic, given that the ram pressure at the leading edge of the comet is directly proportional to atmospheric density, leading to the onset of fragmentation at higher altitudes. As before (see \S\ref{sec:discussion_comet_density}), the rate of deformation, and growth of hydrodynamic instabilities will increase significantly in dense planetary atmospheres. There will therefore be a particularly rapid deposition of mass and energy into the surrounding atmosphere. Example cometary trajectories, demonstrating the effective aerobraking, and fragmentation of $500-1000\,$m radius comets in a 10\,bar atmosphere are included in appendix~\ref{appendix:comet_trajectories_dense_atmos} for reference.

\section{Discussion}
\label{sec:discussion}

We have seen in \S\ref{sec:results} that the dynamical ram pressure experienced by comets during atmospheric entry will always exceed their (low) tensile strength at high altitudes. Yet, there is a large diversity in their subsequent interaction with the atmosphere. As predicted by simple physical arguments (see equation~\ref{eq:qual_rcrit}), this interaction, and therefore the fate of comets in the atmosphere, is determined by three key parameters: (i) the comet's initial radius, (ii) the comet's bulk density, and (iii) the atmospheric surface density. 

We demonstrate that there exists a sharp transition, at a radius of about 150\,m, separating the survival of larger comets, which reach the surface essentially intact, and smaller comets, which experience rapid, successive, and ultimately catastrophic fragmentation at high altitudes. These small comets deposit essentially all of their initial mass, and kinetic energy into the surrounding atmosphere within a fraction of a scale height. This transition, characterising the fate of comets during atmospheric entry, is shown to be independent of initial velocity. These results are therefore independent of cometary dynamics in the planetary system before impact, and require only knowledge of a planet's atmospheric profile.

The location of this transition is very sensitive to both the bulk density of cometary nuclei and the atmospheric surface density. For particularly low density (highly porous) comets, and/or large atmospheric surface densities, we demonstrate that comets must be larger than 1\,km to avoid catastrophic fragmentation, and near-total mass ablation. Observational evidence suggests that (Solar system) comets span a relatively wide-range of bulk densities \citep{Kokotanekova2017, Groussin2019}, and that there was a quick transition in Earth's atmospheric surface pressure, from $\sim\,$100\,bar \citep[following the degassing of Earth's magma ocean;][]{Elkins-Tanton2008}, to less than 1\,bar by the start of the Archean \citep[albeit these constraints remain subject to debate]{CatlingZahnle2020}. The flux of small comets incident at Earth's surface will therefore have changed as a function of time, with this critical size decreasing from roughly $1\,$km to $100\,$m during the first 500\,Myr post Moon-formation.

We discuss in \S\ref{sec:discussion_preb_chem} the implications of these results for the cometary delivery of prebiotic feedstock molecules, and comment in \S\ref{sec:discussion_atmospheres} on how the detailed modelling of cometary fragmentation during atmospheric entry is necessary to support the accurate modelling of the atmospheric response to cometary impacts. In \S\ref{sec:thermal_evolution} we discuss the thermal evolution of comets during atmospheric entry, before detailing model limitations in \S\ref{sec:discussion_limitations}.

\subsection{Implications for prebiotic chemistry}
\label{sec:discussion_preb_chem}

Cometary impacts have been invoked by several authors as a promising mechanism to deliver simple prebiotic feedstock molecules, such as hydrogen cyanide (HCN), to local environments in high concentrations \citep[e.g.,][]{Oro1961, Clark1988, Chyba1992, ToddOberg2020, Zellner2020}. The delivery of simple feedstock molecules is however, at most, a necessary but not sufficient criteria for plausible origins-of-life scenarios. 

A particular challenge facing the use of HCN in prebiotic chemistry is that it will undergo hydrolysis in aqueous solution (to formamide, and then formic acid) at a rate dependent on the temperature of its local environment \citep{Miyakawa2002, Todd2024}. It is likely this will occur rapidly, given the relative abundance of water concurrently delivered during cometary impacts. For this reason, many scenarios invoke the stockpiling of cyanide \citep[e.g., as ferrocyanide salts;][]{Sasselov2020}, as well as a subsequent heat source, which is required to eventually liberate free cyanide. This heat source has been attributed to further impacts or geothermal heating \citep{Patel2015}, although we note that the probability of a subsequent impact in the same location is extremely low, and these scenarios therefore require a very high cometary impact rate \citep{Anslow2024}. 

Alternatively, given that the hydrolysis rate of HCN is substantially decreased at low temperatures \citep[with eutectic freezing suggested to permit the polymerisation of HCN;][]{Miyakawa2002}, cometary impacts onto ice sheets on the early Earth may provide an alternative pathway for successful prebiotic chemistry. The glaciogenic accumulation of cosmic dust has recently been shown to reach prebiotically relevant concentrations \citep{Walton2024}, thereby providing an independent source of other key feedstock molecules for cyanosulfidic prebiotic chemistry. This scenario would of course require the presence of glaciers on the early Earth, which is presently unclear. We note, however, that there exists evidence for glaciers as early as 2.5\,Ga \citep{Kirschvink2000}, and that models supporting a cold Hadean Earth \citep[e.g.,][]{Kadoya2020} independently invoke high impact fluxes.

The probability of further prebiotic chemical reactions required for the synthesis of the building blocks of life, succeeding and occurring in the correct sequence is, in any case, extremely low \citep[e.g.,][]{Rimmer2023}. Thus, for any prebiotic chemical scenarios requiring an initial cometary impact, whether for the stockpiling of (intermediate) ferrocyanide salts or the delivery of HCN to geographically localised ice sheets, it is imperative to maximise the number of cometary impacts able to deliver key feedstock molecules.

Previous work has highlighted a significant challenge facing the cometary delivery of simple prebiotic feedstock molecules, limiting the number of suitable cometary impacts: the extremely high pressures, and temperatures experienced during impacts \citep{PierazzoChyba1999, ToddOberg2020}. \citet{ToddOberg2020} demonstrate that the survival of HCN decreases exponentially with both comet radius, and impact velocity, restricting cometary delivery to small comets (at most, km-scale), at low impact velocities (ideally $\lesssim\,$$15\,{\rm km\,s}^{-1}$). Here, we show that comets smaller than about $100\,$m are also unsuitable for cometary delivery, given they will unavoidably, and catastrophically fragment at high altitudes, depositing the entirety of their initial mass (and energy) into the surrounding atmosphere\footnote{Whilst it is possible that the deposition of large amounts of energy, and volatile-rich material into the atmosphere may promote the formation of various prebiotic feedstock species, this interaction will depend sensitively on atmospheric composition. Moreover, this is also possible during asteroidal impacts \citep{Parkos2018}, which will be significantly more numerous.}.

The implications of this additional constraint are potentially severe. The size-frequency distribution of Jupiter-family comets follows a power-law, with differential slope of around $-2.9$ \citep{Snodgrass2011, Fernandez2013}, such that the number of cometary impacts on Earth are dominated, in number, by the smallest comets in the distribution\footnote{For context, this means that there will be roughly a thousand 100\,m comets for every 1\,km comet impacting the Earth.}. These comets will not survive atmospheric entry, and therefore cannot deliver any prebiotic feedstock molecules to local environments. The survival of key feedstock molecules during cometary impacts in large comets is therefore particularly important, which is investigated further in \citet{McDonald2025}. In any case, given that cometary impacts are from predominantly small bodies unable to survive atmospheric entry, it is likely that only a very small subset of comets will successfully deliver HCN to the early Earth. Quantitative investigation into the subset of comets able to deliver HCN to the early Earth is reserved for an upcoming study (Anslow \& McDonald et al., in prep.).

The atmospheric filtering of small comets is particularly challenging for planets with thick, high mean molecular weight atmospheres, given the minimum cometary diameter required to survive atmospheric entry rapidly approaches the maximum diameter able to deliver key feedstock molecules. This is likely the case for Venus, with atmospheric surface pressure approaching 100\,bar, with only comets larger than about 1\,km able to survive atmospheric entry. Conversely, this is much less significant for Mars, given its very tenuous atmosphere is unable to prevent even very small bodies ($\sim\,$10\,m) from reaching the surface intact. If cometary delivery is indeed an important step necessary for successful prebiotic chemistry, this therefore lends further support to the possibility that life could have emerged on Mars, providing further motivation for Mars Sample Return. Not only this, but high precision measurement of returned organics would also be possible, potentially allowing for the identification of isotopically distinct \citep[][]{Marty2017} cometary material from the Martian surface.

\subsubsection{Is the fragmentation of small comets beneficial for prebiotic chemistry?}

Here, we consider in more detail the delivery of prebiotic feedstock molecules from comets that fragment at low altitude above the surface (e.g., $R_0=150\,$m; figure~\ref{fig:comet_trajectory_gallery}). There are several arguments to suggest this may in fact improve the survival of feedstock molecules during hypervelocity impact, which we discuss here. 

First, these fragments will be significantly deformed, and therefore experience a massive aerodynamic drag force during the final stages of their descent. As seen in figure~\ref{fig:zoom_in_panel}, these fragments reach the surface with only a small fraction of the comet's initial velocity. Given the survival of prebiotic feedstock molecules, such as HCN, increases exponentially for lower impact velocities \citep{ToddOberg2020}, these fragments will deliver a much larger fraction of their prebiotic feedstock inventory than a comet would otherwise deliver, in the absence of an atmosphere.

Second, due to both the fragmentation of the comet, and rapid mass ablation post-fragmentation, each child fragment will be several times less massive than the parent comet. Again, the survival of feedstock molecules increases exponentially with decreasing size \citep{ToddOberg2020}, which further suggests these fragments will deliver prebiotic feedstock molecules more effectively than the comet would in the absence of any atmosphere.

Third, these child fragments move apart from each other during the final stages of atmospheric descent, due to the interaction of their bow shocks. Depending on the magnitude of their transverse acceleration, these fragments will impact the surface downrange from each other, forming a small cluster of craters. Whilst the number, and size of these impact craters will vary, some concentration of prebiotic feedstock molecules will be delivered to a much larger number of environments, than in the absence of an atmosphere. One comet will never deliver every feedstock molecule needed for prebiotic chemistry; indeed, many reactions must occur at different times, and will each require a different inventory of feedstock molecules \citep[e.g.,][]{Rimmer2023}. A cluster of local environments, each containing key prebiotic feedstock molecules may therefore provide a more promising setting for prebiotic chemistry, compared to a single large crater formed by a large km-sized comet.

There is, however, a potentially important caveat to this discussion. During fragmentation, the child fragments separate, and at some point develop individual bow shocks \citep[e.g.,][]{PasseyMelosh1980} in what is an extremely complex interaction with the atmosphere. The bow shock initially surrounding the comet is an important physical barrier, shielding its leading edge from direct interaction with the atmospheric flow \citep[see also \S\ref{sec:thermal_evolution}]{Silber2018}. If during this process the fragments interact, for some time, with the atmospheric flow, this might accelerate the transfer of heat into the comet's interior. The temperature of the shock-heated atmospheric gases is approximately $10^4\,$K \citep[e.g.,][]{Johnston2018}, comparable to (or even greater than) the peak temperatures reached during impact \citep{PierazzoChyba1999}. This could therefore destroy a significant fraction of the comet's initial prebiotic feedstock inventory pre-impact.

\subsection{Mass and energy deposition in the atmosphere}
\label{sec:discussion_atmospheres}

Cometary impacts have been long considered a potential explanation for Jupiter's super-Solar metallicity \citep{Mahaffy2000, MullerHelled2024}, an idea that has recently been extended to young hot Jupiters \citep{Zhang2023, Tsai2023}. This is supported, for hot Jupiters, by detailed climate modelling of the atmospheric response to cometary impacts \citep{SainsburyMartinezWalsh2024}, with this modelling effort recently extended to include terrestrial exoplanets \citep{Sainsbury-Martinez2024}. The atmospheric response of terrestrial exoplanets to cometary impacts is found to be, potentially, both observable and extended, driven by the deposition of water in the upper atmosphere. This water acts as an additional source of opacity at high altitudes, which causes the heating of the upper atmosphere, and the cooling of the lower atmosphere \citep[by as much as $5\,$K;][]{Sainsbury-Martinez2024}. Importantly, the atmospheric response will be significantly reduced if the deposition of water is instead concentrated in the dense lower atmosphere.

The extent, and location of mass and energy deposition into the atmosphere during cometary impacts is therefore crucially important in order to accurately characterise the atmospheric response to cometary impacts. We demonstrate in \S\ref{sec:results} that comets of all sizes deposit mass and energy into the atmosphere, varying significantly (in location and extent) with cometary diameter (see figure~\ref{fig:comet_trajectory_gallery}), predominantly occurring during fragmentation. Small comets that catastrophically fragment at high altitudes deposit almost the entirety of their initial mass, and energy within a fraction of a scale height. The altitude at which this occurs is determined by their initial radius (see figure~\ref{fig:comet_trajectory_gallery}), given this controls their rate of deformation, and fragmentation in the atmosphere. Large comets, on the other hand, do not fragment, and lose only a very small fraction of their initial mass due to ablation, predominantly in the dense lower atmosphere. Observations of small meteoroids in Earth's atmosphere support this conclusion that fragmentation plays a much more important role than both aerobraking and mass ablation \citep{Popova2019}. 

Uncertainty in mass deposition profiles still remains however for large comets that do not fragment. Considering only the upper atmosphere (most sensitive to the additional opacity source provided by cometary water), before the ram pressure exceeds the comet's tensile strength, mass deposition is driven only by ablation (see figure~\ref{fig:comet_trajectory_gallery}). The strength, and duration of any atmospheric response is therefore very sensitive to the comet's heat transfer coefficient (see \S\ref{sec:methods_mass_ablation}), which can only be determined via detailed numerical modelling \citep[e.g.,][]{Johnston2018}. We summarise existing literature, describing the relatively poor constraints on this coefficient for km-scale impactors in \S\ref{sec:methods_mass_ablation}, which will unavoidably be inherited by any mass deposition profiles adopted in these studies.

In summary, mass and energy deposition in the atmosphere is dominated by fragmentation rather than mass ablation, and is therefore extremely sensitive to initial radius, with only small comets ($\lesssim\,100\,$m) depositing a significant fraction of their initial mass (energy) into the surrounding atmosphere. This occurs rapidly in what is a runaway process whereby comets spread their mass over an ever-increasing area, further increasing atmospheric drag, and accelerating their deformation. For these comets, mass loss occurs within a fraction of a scale height, at an altitude determined by their initial radius. In contrast, larger comets lose only a very small fraction of their initial mass due to ablation, which occurs predominantly in the dense lower atmosphere. The atmospheric response to cometary impacts is therefore likely overestimated when assuming significant mass loss from large comets in the upper atmosphere.
The numerical model described in this study, available as a python package\footnote{\href{https://github.com/richard17a/atmosentry}{https://github.com/richard17a/atmosentry}.}, provides this key input required by the computationally expensive 3D climate models used to characterise the atmospheric response to cometary impacts: mass, and energy deposition profiles (as a function of altitude) from comets of any size. 

\subsection{The thermal evolution of comets}
\label{sec:thermal_evolution}

Throughout this study we do not consider the thermal evolution of comets during atmospheric entry. Given however the friable and highly porous nature of cometary nuclei, their interaction with the atmosphere may, in principle, drive significant thermal processing. In the context of prebiotic chemistry, this would have substantial negative consequences, and we therefore revisit this assumption below.

The nature of the interaction between a comet, and the atmosphere is crucially important when determining the mechanism, and efficiency of heat transfer. Observations reveal that this interaction changes significantly with altitude \citep[e.g.,][]{Biberman1980}, with high-fidelity simulations (coupling fluid dynamics with radiative transfer) required to accurately model the heating of meteoroids during atmospheric entry, and constrain key model parameters \citep[e.g.,][]{Johnston2018}. 
To a good approximation, however, the nature of the local flow regime can be estimated via the Knudsen number \citep{Silber2018, Popova2019},
\begin{equation}
    {\rm Kn} = \dfrac{l}{r} = 0.1 \left(\dfrac{l}{10\,{\rm cm}}\right)\left(\dfrac{r}{1\,{\rm m}}\right)^{-1},
\end{equation}
which is defined as the ratio of the atmospheric mean free path ($l$), to a characteristic length-scale of the body (i.e., its radius $r$). Values of ${\rm Kn}$ below $0.1$ correspond to the continuous flow regime. This is satisfied even for metre-sized comets in the rarefied upper atmosphere, where the mean free path is large \citep[$\sim\,$$10\,$cm;][]{Campbell-BrownKoschny2004}.

In the continuous flow regime a shock front (with density several orders of magnitude greater than the surrounding atmosphere) forms at the leading edge of the comet, preventing the direct interaction of hot atmospheric gases with the comet. This protection is crucially important for the intact survival of large comets during atmospheric entry. Comets, protected by this shock front, lose mass due to ablation in the continuous flow regime driven by the absorption of a large radiative flux (emitted by the shock front) at their leading edge \citep[e.g.,][]{Johnston2018, Silber2018}.

It is this radiative flux from the shock front that will drive thermal evolution during atmospheric entry, heating the leading edge of comets to very high temperatures. While this is evidenced by the fusion crust found on meteorites \citep[e.g.,][]{GengeGrady1999}, the extent of thermal evolution beyond a comet's leading edge will depend on the transfer of heat throughout its body.
The degree to which a body can be considered isothermal is characterised by the dimensionless Biot number, ${\rm Bi}$, given by
\begin{equation}
    {\rm Bi} = \dfrac{h L }{k} = 10^3\left(\dfrac{h}{10^4\,{\rm W\,m^{-2}\,K}}\right) \left(\dfrac{k}{1\,{\rm W\,m^{-1}\,K^{-1}}}\right)^{-1} \left(\dfrac{L}{1\,{\rm m}}\right),
\end{equation}
where $h$ is the heat transfer coefficient, $k$ the meteoroid's thermal conductivity, and $L$ a characteristic length scale. For ${\rm Bi}>10$, it is not reasonable to assume an isothermal temperature distribution, and there will exist large temperature gradients within the comet.

The appropriate heat transfer coefficient is not immediately obvious, but is likely dominated by the radiative flux received from the atmospheric shock front. It has units of power per area per Kelvin, and we assume $h\sim 10^4\,{\rm W\,m^{-2}\,K}$ \citep[using the results from][]{Johnston2018}. The thermal conductivity of ice is low, $k \sim 1\,{\rm W\,m^{-1}\,K^{-1}}$ \citep{Aschwanden2012}, and thus only very small ($\sim\,{\rm \mu m}$) meteorites will be heated throughout to high temperature. Large comets, able to reach the surface intact (i.e., $\gg$\,$1\,$m) will not be heated throughout. Rather, as we show next, only the very outer layers undergo any significant heating during atmospheric entry.

Whilst the leading edge of the comet will be heated to high temperatures during atmospheric entry, the significant heating of its interior requires the very rapid transfer of heat due to the short timescale on which comets pass through atmosphere (on the order of 10\,s). Assuming spherical symmetry, the heat equation governing the comet's thermal evolution during atmospheric entry (in response to the shock-front at its leading edge) is then
\begin{equation*}
    \dfrac{1}{\alpha} \dfrac{\partial T(r, t)}{\partial t} = \dfrac{1}{r^2} \dfrac{\partial}{\partial r}\left(r^2 \dfrac{\partial T(r, t)}{\partial r}\right).
\end{equation*}
To leading order, the radial extent of heat transfer over a time $\delta t \sim 10\,$s (i.e., the skin-depth) is
\begin{equation*}
    \delta r \sim \sqrt{\alpha \delta t} \sim 0.1 \left(\dfrac{\alpha}{10^{-8}\,{\rm m^2\,s}^{-1}}\right)^{1/2}\left(\dfrac{\delta t}{10\,{\rm s}}\right)^{1/2}\,{\rm mm}.
\end{equation*}
Despite a lack of knowledge regarding key physical properties of cometary nuclei, the thermal diffusivity of ice is known to be low \citep[$\sim\,$$10^{-7}\,{\rm m^2\,s}^{-1}$;][]{Fukusako1990}, and found to increase with bulk density \citep{Fukusako1990}. Thus, given the low density of comets \citep[as evidenced by comet 67P;][]{Jorda2016}, it is reasonable to suggest that the thermal diffusivity of comets might be significantly lower than of ice. We assume a thermal diffusivity $\alpha\sim10^{-8}\,{\rm m^2\,s}^{-1}$ following \citet{PrialnikJewitt2022}. Only the outer $\sim\,$$0.1\,$mm of comets will be heated to temperatures of the order $10^4\,$K.

Finally, we note that any thermal degradation will be largely confined to the comet's leading edge. The survival of any prebiotic feedstock molecules near the leading edge is already very unlikely, given this region is heated to the highest temperatures during impact with the surface \citep{PierazzoChyba1999}. Thus, the thermal degradation of prebiotic feedstock molecules will be insignificant for all comets able to reach the surface, avoiding fragmentation in the atmosphere (see \S\ref{sec:results}).

\subsection{Model limitations}
\label{sec:discussion_limitations}

In this work we present a (deliberately) simple numerical model for the atmospheric entry of cometary impactors. We adopt several semi-analytical parameterisations to describe the ablation, deformation, and fragmentation of comets as they interact with the atmosphere. Many of the complex physical processes that occur during atmospheric entry are therefore not accurately described by this model. At this expense, however, we are able to clearly motivate our results in terms of simple physical arguments, and the model retains sufficient computational simplicity to cover a wide range of parameter space (both comet, and atmosphere). Despite this, our model is in good agreement with existing studies (see appendix~\ref{appendix:chyba_validation}), and is based closely on the model presented in \citet{KorycanskyZahnle2005}, which is successfully validated against Venus' crater population. 

We adopt a continuous fragmentation (pancake) model in this study, however as discussed in \S\ref{sec:deformation_fragmentation} there also exist several discrete fragmentation models that can also be used to model the fragmentation of meteoroids in the atmosphere. A detailed asteroid fragmentation model comparison was presented in \citet{Register2017}, including both continuous and discrete models. They found, using constraints from the Chelyabinsk airburst, that energy deposition in the atmosphere is driven by small fragments behaving in a weakly cohesive, aggregate fashion. Discrete fragmentation models only succeeded when closely mimicking a cloud-like behaviour \citep{Register2017}. Thus, given the significantly lower material strength (and generally weakly cohesive nature) of comets, it is very unlikely that discrete fragmentation models will accurately describe their interaction with the atmosphere.

Finally, for the cometary delivery of prebiotic feedstock molecules, a main limitation of this work is that we do not model the thermal evolution of comets during atmospheric entry. Whilst this is consistent with existing literature, the high porosity of comets potentially renders them particularly vulnerable to thermal processing. We demonstrate in \S\ref{sec:thermal_evolution} that the thermal evolution of large comets will be negligible, given they are effectively shielded from direct interaction with the atmospheric flow by the shock front at their leading edge \citep[e.g.,][]{Silber2018, Popova2019}. However, the same is not necessarily true for smaller comets that fragment close to the surface. As discussed in \S\ref{sec:methods}, during fragmentation the collective bow shock disrupts, potentially allowing for direct interaction between the fragmenting comet and hot atmospheric gases. If during this processes hot gases are driven into the child fragments, this will rapidly accelerate the transfer of heat into their interiors, potentially reaching temperatures as high as $10^4\,$K on very short timescales. A significant fraction of a comet's prebiotic inventory would therefore be destroyed before impact, further restricting the subset of comets able to deliver prebiotic feedstock molecules to only large ($\sim\,500\,$m) comets that survive atmospheric entry intact. This is, however, a complex comet-atmosphere interaction that is beyond the scope of this study, and we therefore stress this important caveat when considering cometary fragments in prebiotic chemical scenarios.

\section{Conclusions}
\label{sec:conclusion}

We have developed a simple, open-source, numerical package to model the impact of comets into planetary atmospheres. The model includes specific semi-analytical parameterisations for the ablation, deformation, and fragmentation of comets during atmospheric entry. The model retains sufficient computational simplicity to efficiently investigate a wide range of parameter space, and allows us to clearly motivate the key physical processes that control the interaction of comets with the atmosphere. 

We first apply our model to the cometary delivery of prebiotic feedstock molecules, which requires the survival of comets (or fragments thereof) to the surface, for the concentration of key feedstock molecules in localised environments. The survival of comets is determined by three parameters: (i) the comet's initial radius, (ii) the comet's bulk density, and (iii) the atmospheric surface density. We find there exists a very sharp transition between the survival, and catastrophic fragmentation of comets in the atmosphere. For Earth-like atmospheres, this transition occurs at a radius of $\sim\,$150\,m, below which comets deform rapidly, losing both their mass and kinetic energy to the surrounding atmosphere. This critical radius increases with increasing atmospheric surface density, and decreasing cometary density, often exceeding $1\,$km.

Finally, we demonstrate that the deposition of mass and kinetic energy in planetary atmospheres is similarly sensitive to comets' initial radius. Mass and energy loss in the atmosphere is dominated by fragmentation, which is found to have a dominant effect in comparison to mass ablation. The deformation, and fragmentation of small comets spreads their mass over a large area, further increasing atmospheric drag, and driving enhanced deformation. This feedback causes the rapid deposition of mass within a fraction of an atmospheric scale height, at an altitude determined by the comet's initial radius. Large comets, on the other hand, do not fragment in the atmosphere, and lose only a small fraction of their initial mass to ablation in the dense lower atmosphere.
The location, and extent of mass loss determines the strength, and duration of any atmospheric response to cometary impacts, which is accordingly extremely sensitive to comets' initial radius, and the detailed modelling of cometary fragmentation.

\section*{Acknowledgements}

We thank an anonymous reviewer for their constructive comments that have helped improve the quality of this manuscript.
We thank Felix Sainsbury-Martinez for valuable discussions regarding the response of planetary atmospheres to cometary impacts.
R.J.A. acknowledges the Science and Technology Facilities Council (STFC) for a
PhD studentship. A.B. acknowledges the support of a Royal Society University Research Fellowship, URF/R1/211421. 
C.H.M. gratefully acknowledges the Leverhulme Centre for Life in the Universe at the University of Cambridge for support through Joint Collaborations Research Project Grant GAG/382.

\section*{Data Availability}

The data and code used in this work are publicly available at \url{https://github.com/richard17a/atmosentry}.



\bibliographystyle{mnras}
\bibliography{main} 

\begin{thebibliography}{}
\makeatletter
\relax
\def\mn@urlcharsother{\let\do\@makeother \do\$\do\&\do\#\do\^\do\_\do\%\do\~}
\def\mn@doi{\begingroup\mn@urlcharsother \@ifnextchar [ {\mn@doi@} {\mn@doi@[]}}
\def\mn@doi@[#1]#2{\def\@tempa{#1}\ifx\@tempa\@empty \href {http://dx.doi.org/#2} {doi:#2}\else \href {http://dx.doi.org/#2} {#1}\fi \endgroup}
\def\mn@eprint#1#2{\mn@eprint@#1:#2::\@nil}
\def\mn@eprint@arXiv#1{\href {http://arxiv.org/abs/#1} {{\tt arXiv:#1}}}
\def\mn@eprint@dblp#1{\href {http://dblp.uni-trier.de/rec/bibtex/#1.xml} {dblp:#1}}
\def\mn@eprint@#1:#2:#3:#4\@nil{\def\@tempa {#1}\def\@tempb {#2}\def\@tempc {#3}\ifx \@tempc \@empty \let \@tempc \@tempb \let \@tempb \@tempa \fi \ifx \@tempb \@empty \def\@tempb {arXiv}\fi \@ifundefined {mn@eprint@\@tempb}{\@tempb:\@tempc}{\expandafter \expandafter \csname mn@eprint@\@tempb\endcsname \expandafter{\@tempc}}}

\bibitem[\protect\citeauthoryear{{Ahrens}, {Takata}, {O'Keefe}  \& {Orton}}{{Ahrens} et~al.}{1994}]{Ahrens1994}
{Ahrens} T.~J.,  {Takata} T.,  {O'Keefe} J.~D.,   {Orton} G.~S.,  1994, \mn@doi [\grl] {10.1029/94GL01325}, \href {https://ui.adsabs.harvard.edu/abs/1994GeoRL..21.1087A} {21, 1087}

\bibitem[\protect\citeauthoryear{{Anslow}, {Bonsor}  \& {Rimmer}}{{Anslow} et~al.}{2023}]{Anslow2023}
{Anslow} R.~J.,  {Bonsor} A.,   {Rimmer} P.~B.,  2023, \mn@doi [Proceedings of the Royal Society of London Series A] {10.1098/rspa.2023.0434}, \href {https://ui.adsabs.harvard.edu/abs/2023RSPSA.47930434A} {479, 20230434}

\bibitem[\protect\citeauthoryear{{Anslow}, {Bonsor}, {Rimmer}, {Rae}, {McDonald}  \& {Walton}}{{Anslow} et~al.}{2025}]{Anslow2024}
{Anslow} R.~J.,  {Bonsor} A.,  {Rimmer} P.~B.,  {Rae} A.~S.~P.,  {McDonald} C.~H.,   {Walton} C.~R.,  2025, \mn@doi [Proceedings of the Royal Society of London Series A] {10.1098/rspa.2024.0327}, \href {https://ui.adsabs.harvard.edu/abs/2025RSPSA.48140327A} {481, 20240327}

\bibitem[\protect\citeauthoryear{{Artemieva} \& {Shuvalov}}{{Artemieva} \& {Shuvalov}}{2001}]{ArtemievaShuvalov2001}
{Artemieva} N.~A.,  {Shuvalov} V.~V.,  2001, \mn@doi [\jgr] {10.1029/2000JE001264}, \href {https://ui.adsabs.harvard.edu/abs/2001JGR...106.3297A} {106, 3297}

\bibitem[\protect\citeauthoryear{Aschwanden, Bueler, Khroulev  \& Blatter}{Aschwanden et~al.}{2012}]{Aschwanden2012}
Aschwanden A.,  Bueler E.,  Khroulev C.,   Blatter H.,  2012, \mn@doi [Journal of Glaciology] {10.3189/2012JoG11J088}, 58, 441–457

\bibitem[\protect\citeauthoryear{{Attree} et~al.,}{{Attree} et~al.}{2018}]{Attree2018}
{Attree} N.,  et~al., 2018, \mn@doi [\aap] {10.1051/0004-6361/201732155}, \href {https://ui.adsabs.harvard.edu/abs/2018A&A...611A..33A} {611, A33}

\bibitem[\protect\citeauthoryear{{Baldwin} \& {Sheaffer}}{{Baldwin} \& {Sheaffer}}{1971}]{BaldwinSheaffer1971}
{Baldwin} B.,  {Sheaffer} Y.,  1971, \mn@doi [\jgr] {10.1029/JA076i019p04653}, \href {https://ui.adsabs.harvard.edu/abs/1971JGR....76.4653B} {76, 4653}

\bibitem[\protect\citeauthoryear{{Becker} et~al.,}{{Becker} et~al.}{2019}]{Becker2019}
{Becker} S.,  et~al., 2019, \mn@doi [Science] {10.1126/science.aax2747}, \href {https://ui.adsabs.harvard.edu/abs/2019Sci...366...76B} {366, 76}

\bibitem[\protect\citeauthoryear{{Benz} \& {Asphaug}}{{Benz} \& {Asphaug}}{1994}]{BenzAsphaug1994}
{Benz} W.,  {Asphaug} E.,  1994, in Lunar and Planetary Science Conference. Lunar and Planetary Science Conference.
p.~101

\bibitem[\protect\citeauthoryear{{Biberman}, {Bronin}  \& {Brykin}}{{Biberman} et~al.}{1980}]{Biberman1980}
{Biberman} L.~M.,  {Bronin} S.~Y.,   {Brykin} M.~V.,  1980, \mn@doi [Acta Astronautica] {10.1016/0094-5765(80)90116-2}, \href {https://ui.adsabs.harvard.edu/abs/1980AcAau...7...53B} {7, 53}

\bibitem[\protect\citeauthoryear{{Borovi{\v{c}}ka} \& {Spurn{\'y}}}{{Borovi{\v{c}}ka} \& {Spurn{\'y}}}{1996}]{BorovickaSpurny1996}
{Borovi{\v{c}}ka} J.,  {Spurn{\'y}} P.,  1996, \mn@doi [\icarus] {10.1006/icar.1996.0104}, \href {https://ui.adsabs.harvard.edu/abs/1996Icar..121..484B} {121, 484}

\bibitem[\protect\citeauthoryear{{Boslough}, {Crawford}, {Robinson}  \& {Trucano}}{{Boslough} et~al.}{1994}]{Boslough1994}
{Boslough} M.~B.,  {Crawford} D.~A.,  {Robinson} A.~C.,   {Trucano} T.~G.,  1994, \mn@doi [\grl] {10.1029/94GL01582}, \href {https://ui.adsabs.harvard.edu/abs/1994GeoRL..21.1555B} {21, 1555}

\bibitem[\protect\citeauthoryear{{Bronshten}}{{Bronshten}}{1983}]{Bronshten1983}
{Bronshten} V.~A.,  1983, {Physics of Meteoric Phenomena}.
Springer Dordrecht

\bibitem[\protect\citeauthoryear{{Campbell-Brown} \& {Koschny}}{{Campbell-Brown} \& {Koschny}}{2004}]{Campbell-BrownKoschny2004}
{Campbell-Brown} M.~D.,  {Koschny} D.,  2004, \mn@doi [\aap] {10.1051/0004-6361:20041001-1}, \href {https://ui.adsabs.harvard.edu/abs/2004A&A...418..751C} {418, 751}

\bibitem[\protect\citeauthoryear{{Catling} \& {Zahnle}}{{Catling} \& {Zahnle}}{2020}]{CatlingZahnle2020}
{Catling} D.~C.,  {Zahnle} K.~J.,  2020, \mn@doi [Science Advances] {10.1126/sciadv.aax1420}, \href {https://ui.adsabs.harvard.edu/abs/2020SciA....6.1420C} {6, eaax1420}

\bibitem[\protect\citeauthoryear{{Chyba}}{{Chyba}}{1990}]{Chyba1990}
{Chyba} C.~F.,  1990, \mn@doi [\nat] {10.1038/343129a0}, \href {https://ui.adsabs.harvard.edu/abs/1990Natur.343..129C} {343, 129}

\bibitem[\protect\citeauthoryear{{Chyba} \& {Sagan}}{{Chyba} \& {Sagan}}{1992}]{Chyba1992}
{Chyba} C.,  {Sagan} C.,  1992, \mn@doi [\nat] {10.1038/355125a0}, \href {https://ui.adsabs.harvard.edu/abs/1992Natur.355..125C} {355, 125}

\bibitem[\protect\citeauthoryear{{Chyba}, {Thomas}  \& {Zahnle}}{{Chyba} et~al.}{1993}]{Chyba1993}
{Chyba} C.~F.,  {Thomas} P.~J.,   {Zahnle} K.~J.,  1993, \mn@doi [\nat] {10.1038/361040a0}, \href {https://ui.adsabs.harvard.edu/abs/1993Natur.361...40C} {361, 40}

\bibitem[\protect\citeauthoryear{{Clark}}{{Clark}}{1988}]{Clark1988}
{Clark} B.~C.,  1988, \mn@doi [Origins of Life] {10.1007/BF01804671}, \href {https://ui.adsabs.harvard.edu/abs/1988OrLi...18..209C} {18, 209}

\bibitem[\protect\citeauthoryear{{Collins}, {Melosh}  \& {Marcus}}{{Collins} et~al.}{2005}]{Collins2005}
{Collins} G.~S.,  {Melosh} H.~J.,   {Marcus} R.~A.,  2005, \mn@doi [Meteoritics \& Planetary Science] {10.1111/j.1945-5100.2005.tb00157.x}, \href {https://ui.adsabs.harvard.edu/abs/2005M&PS...40..817C} {40, 817}

\bibitem[\protect\citeauthoryear{{Collins} et~al.,}{{Collins} et~al.}{2022}]{Collins2022}
{Collins} G.~S.,  et~al., 2022, \mn@doi [Journal of Geophysical Research (Planets)] {10.1029/2021JE007149}, \href {https://ui.adsabs.harvard.edu/abs/2022JGRE..12707149C} {127, e07149}

\bibitem[\protect\citeauthoryear{{Crawford}}{{Crawford}}{1997}]{Crawford1997}
{Crawford} D.~A.,  1997, \mn@doi [Annals of the New York Academy of Sciences] {10.1111/j.1749-6632.1997.tb48340.x}, \href {https://ui.adsabs.harvard.edu/abs/1997NYASA.822..155C} {822, 155}

\bibitem[\protect\citeauthoryear{{Ehrenfreund} et~al.,}{{Ehrenfreund} et~al.}{2002}]{Ehrenfreund2002}
{Ehrenfreund} P.,  et~al., 2002, in {Lacoste} H.,  ed.,  ESA Special Publication Vol. 518, Exo-Astrobiology. pp 9--14

\bibitem[\protect\citeauthoryear{{Elkins-Tanton}}{{Elkins-Tanton}}{2008}]{Elkins-Tanton2008}
{Elkins-Tanton} L.~T.,  2008, \mn@doi [Earth and Planetary Science Letters] {10.1016/j.epsl.2008.03.062}, \href {https://ui.adsabs.harvard.edu/abs/2008E&PSL.271..181E} {271, 181}

\bibitem[\protect\citeauthoryear{{Fern{\'a}ndez} et~al.,}{{Fern{\'a}ndez} et~al.}{2013}]{Fernandez2013}
{Fern{\'a}ndez} Y.~R.,  et~al., 2013, \mn@doi [\icarus] {10.1016/j.icarus.2013.07.021}, \href {https://ui.adsabs.harvard.edu/abs/2013Icar..226.1138F} {226, 1138}

\bibitem[\protect\citeauthoryear{{Field} \& {Ferrara}}{{Field} \& {Ferrara}}{1995}]{FieldFerrara1995}
{Field} G.~B.,  {Ferrara} A.,  1995, \mn@doi [\apj] {10.1086/175137}, \href {https://ui.adsabs.harvard.edu/abs/1995ApJ...438..957F} {438, 957}

\bibitem[\protect\citeauthoryear{{Flynn}, {Consolmagno}, {Brown}  \& {Macke}}{{Flynn} et~al.}{2018}]{Flynn2018}
{Flynn} G.~J.,  {Consolmagno} G.~J.,  {Brown} P.,   {Macke} R.~J.,  2018, \mn@doi [Chemie der Erde / Geochemistry] {10.1016/j.chemer.2017.04.002}, \href {https://ui.adsabs.harvard.edu/abs/2018ChEG...78..269F} {78, 269}

\bibitem[\protect\citeauthoryear{{Fukusako}}{{Fukusako}}{1990}]{Fukusako1990}
{Fukusako} S.,  1990, \mn@doi [International Journal of Thermophysics] {10.1007/BF01133567}, \href {https://ui.adsabs.harvard.edu/abs/1990IJT....11..353F} {11, 353}

\bibitem[\protect\citeauthoryear{{Genge} \& {Grady}}{{Genge} \& {Grady}}{1999}]{GengeGrady1999}
{Genge} M.~J.,  {Grady} M.~M.,  1999, \mn@doi [Meteoritics \& Planetary Science] {10.1111/j.1945-5100.1999.tb01344.x}, \href {https://ui.adsabs.harvard.edu/abs/1999M&PS...34..341G} {34, 341}

\bibitem[\protect\citeauthoryear{{Groussin} et~al.,}{{Groussin} et~al.}{2019}]{Groussin2019}
{Groussin} O.,  et~al., 2019, \mn@doi [\ssr] {10.1007/s11214-019-0594-x}, \href {https://ui.adsabs.harvard.edu/abs/2019SSRv..215...29G} {215, 29}

\bibitem[\protect\citeauthoryear{{Halliday}}{{Halliday}}{2013}]{Halliday2013}
{Halliday} A.~N.,  2013, \mn@doi [\gca] {10.1016/j.gca.2012.11.015}, \href {https://ui.adsabs.harvard.edu/abs/2013GeCoA.105..146H} {105, 146}

\bibitem[\protect\citeauthoryear{{Herrick} \& {Phillips}}{{Herrick} \& {Phillips}}{1994}]{HerrickPhillips1994}
{Herrick} R.~R.,  {Phillips} R.~J.,  1994, \mn@doi [\icarus] {10.1006/icar.1994.1180}, \href {https://ui.adsabs.harvard.edu/abs/1994Icar..112..253H} {112, 253}

\bibitem[\protect\citeauthoryear{{Hills} \& {Goda}}{{Hills} \& {Goda}}{1993}]{HillsGoda1993}
{Hills} J.~G.,  {Goda} M.~P.,  1993, \mn@doi [\aj] {10.1086/116499}, \href {https://ui.adsabs.harvard.edu/abs/1993AJ....105.1114H} {105, 1114}

\bibitem[\protect\citeauthoryear{{Johnston}, {Stern}  \& {Wheeler}}{{Johnston} et~al.}{2018}]{Johnston2018}
{Johnston} C.~O.,  {Stern} E.~C.,   {Wheeler} L.~F.,  2018, \mn@doi [\icarus] {10.1016/j.icarus.2018.02.026}, \href {https://ui.adsabs.harvard.edu/abs/2018Icar..309...25J} {309, 25}

\bibitem[\protect\citeauthoryear{{Jorda} et~al.,}{{Jorda} et~al.}{2016}]{Jorda2016}
{Jorda} L.,  et~al., 2016, \mn@doi [\icarus] {10.1016/j.icarus.2016.05.002}, \href {https://ui.adsabs.harvard.edu/abs/2016Icar..277..257J} {277, 257}

\bibitem[\protect\citeauthoryear{{Kadoya}, {Krissansen-Totton}  \& {Catling}}{{Kadoya} et~al.}{2020}]{Kadoya2020}
{Kadoya} S.,  {Krissansen-Totton} J.,   {Catling} D.~C.,  2020, \mn@doi [Geochemistry, Geophysics, Geosystems] {10.1029/2019GC008734}, \href {https://ui.adsabs.harvard.edu/abs/2020GGG....2108734K} {21, e2019GC008734}

\bibitem[\protect\citeauthoryear{{Kirschvink}, {Gaidos}, {Bertani}, {Beukes}, {Gutzmer}, {Maepa}  \& {Steinberger}}{{Kirschvink} et~al.}{2000}]{Kirschvink2000}
{Kirschvink} J.~L.,  {Gaidos} E.~J.,  {Bertani} L.~E.,  {Beukes} N.~J.,  {Gutzmer} J.,  {Maepa} L.~N.,   {Steinberger} R.~E.,  2000, \mn@doi [Proceedings of the National Academy of Science] {10.1073/pnas.97.4.1400}, \href {https://ui.adsabs.harvard.edu/abs/2000PNAS...97.1400K} {97, 1400}

\bibitem[\protect\citeauthoryear{{Kokotanekova} et~al.,}{{Kokotanekova} et~al.}{2017}]{Kokotanekova2017}
{Kokotanekova} R.,  et~al., 2017, \mn@doi [\mnras] {10.1093/mnras/stx1716}, \href {https://ui.adsabs.harvard.edu/abs/2017MNRAS.471.2974K} {471, 2974}

\bibitem[\protect\citeauthoryear{Korycansky \& Zahnle}{Korycansky \& Zahnle}{2005}]{KorycanskyZahnle2005}
Korycansky D.,  Zahnle K.,  2005, \mn@doi [Planetary and Space Science] {https://doi.org/10.1016/j.pss.2005.03.002}, 53, 695

\bibitem[\protect\citeauthoryear{{Korycansky}, {Zahnle}  \& {Law}}{{Korycansky} et~al.}{2000}]{Korycansky2000}
{Korycansky} D.~G.,  {Zahnle} K.~J.,   {Law} M.-M.~M.,  2000, \mn@doi [\icarus] {10.1006/icar.2000.6426}, \href {https://ui.adsabs.harvard.edu/abs/2000Icar..146..387K} {146, 387}

\bibitem[\protect\citeauthoryear{{Korycansky}, {Zahnle}  \& {Low}}{{Korycansky} et~al.}{2002}]{Korycansky2002}
{Korycansky} D.~G.,  {Zahnle} K.~J.,   {Low} M.-M.~M.,  2002, \mn@doi [\icarus] {10.1006/icar.2002.6795}, \href {https://ui.adsabs.harvard.edu/abs/2002Icar..157....1K} {157, 1}

\bibitem[\protect\citeauthoryear{{Kral}, {Wyatt}, {Triaud}, {Marino}, {Th{\'e}bault}  \& {Shorttle}}{{Kral} et~al.}{2018}]{Kral2018}
{Kral} Q.,  {Wyatt} M.~C.,  {Triaud} A. H.~M.~J.,  {Marino} S.,  {Th{\'e}bault} P.,   {Shorttle} O.,  2018, \mn@doi [\mnras] {10.1093/mnras/sty1677}, \href {https://ui.adsabs.harvard.edu/abs/2018MNRAS.479.2649K} {479, 2649}

\bibitem[\protect\citeauthoryear{{Mahaffy}, {Niemann}, {Alpert}, {Atreya}, {Demick}, {Donahue}, {Harpold}  \& {Owen}}{{Mahaffy} et~al.}{2000}]{Mahaffy2000}
{Mahaffy} P.~R.,  {Niemann} H.~B.,  {Alpert} A.,  {Atreya} S.~K.,  {Demick} J.,  {Donahue} T.~M.,  {Harpold} D.~N.,   {Owen} T.~C.,  2000, \mn@doi [\jgr] {10.1029/1999JE001224}, \href {https://ui.adsabs.harvard.edu/abs/2000JGR...10515061M} {105, 15061}

\bibitem[\protect\citeauthoryear{{Marty} et~al.,}{{Marty} et~al.}{2016}]{Marty2016}
{Marty} B.,  et~al., 2016, \mn@doi [Earth and Planetary Science Letters] {10.1016/j.epsl.2016.02.031}, \href {https://ui.adsabs.harvard.edu/abs/2016E&PSL.441...91M} {441, 91}

\bibitem[\protect\citeauthoryear{Marty et~al.,}{Marty et~al.}{2017}]{Marty2017}
Marty B.,  et~al., 2017, \mn@doi [Science] {10.1126/science.aal3496}, 356, 1069

\bibitem[\protect\citeauthoryear{{McDonald}, {Bonsor}, {Rae}, {Rimmer}, {Anslow}  \& {Todd}}{{McDonald} et~al.}{2025}]{McDonald2025}
{McDonald} C.,  {Bonsor} A.,  {Rae} A.,  {Rimmer} P.,  {Anslow} R.,   {Todd} Z.,  2025, Icarus (submitted)

\bibitem[\protect\citeauthoryear{{Melosh}}{{Melosh}}{1989}]{Melosh1989}
{Melosh} H.~J.,  1989, {Impact Cratering: A Geologic Process}.
Oxford University Press

\bibitem[\protect\citeauthoryear{Miyakawa, James~Cleaves  \& Miller}{Miyakawa et~al.}{2002}]{Miyakawa2002}
Miyakawa S.,  James~Cleaves H.,   Miller S.~L.,  2002, Origins of life and evolution of the biosphere, 32, 195

\bibitem[\protect\citeauthoryear{{M{\"u}ller} \& {Helled}}{{M{\"u}ller} \& {Helled}}{2024}]{MullerHelled2024}
{M{\"u}ller} S.,  {Helled} R.,  2024, \mn@doi [\apj] {10.3847/1538-4357/ad3738}, \href {https://ui.adsabs.harvard.edu/abs/2024ApJ...967....7M} {967, 7}

\bibitem[\protect\citeauthoryear{{Mumma} \& {Charnley}}{{Mumma} \& {Charnley}}{2011}]{MummaCharnley2011}
{Mumma} M.~J.,  {Charnley} S.~B.,  2011, \mn@doi [\araa] {10.1146/annurev-astro-081309-130811}, \href {https://ui.adsabs.harvard.edu/abs/2011ARA&A..49..471M} {49, 471}

\bibitem[\protect\citeauthoryear{{Or{\'o}}}{{Or{\'o}}}{1961}]{Oro1961}
{Or{\'o}} J.,  1961, \mn@doi [\nat] {10.1038/1911193a0}, \href {https://ui.adsabs.harvard.edu/abs/1961Natur.191.1193O} {191, 1193}

\bibitem[\protect\citeauthoryear{{Osinski}, {Cockell}, {Pontefract}  \& {Sapers}}{{Osinski} et~al.}{2020}]{Osinski2020}
{Osinski} G.~R.,  {Cockell} C.~S.,  {Pontefract} A.,   {Sapers} H.~M.,  2020, \mn@doi [Astrobiology] {10.1089/ast.2019.2203}, \href {https://ui.adsabs.harvard.edu/abs/2020AsBio..20.1121O} {20, 1121}

\bibitem[\protect\citeauthoryear{{Parkos}, {Pikus}, {Alexeenko}  \& {Melosh}}{{Parkos} et~al.}{2018}]{Parkos2018}
{Parkos} D.,  {Pikus} A.,  {Alexeenko} A.,   {Melosh} H.~J.,  2018, \mn@doi [Journal of Geophysical Research (Planets)] {10.1002/2017JE005393}, \href {https://ui.adsabs.harvard.edu/abs/2018JGRE..123..892P} {123, 892}

\bibitem[\protect\citeauthoryear{{Passey} \& {Melosh}}{{Passey} \& {Melosh}}{1980}]{PasseyMelosh1980}
{Passey} Q.~R.,  {Melosh} H.~J.,  1980, \mn@doi [\icarus] {10.1016/0019-1035(80)90072-X}, \href {https://ui.adsabs.harvard.edu/abs/1980Icar...42..211P} {42, 211}

\bibitem[\protect\citeauthoryear{{Patel}, {Percivalle}, {Ritson}, {Duffy}  \& {Sutherland}}{{Patel} et~al.}{2015}]{Patel2015}
{Patel} B.~H.,  {Percivalle} C.,  {Ritson} D.~J.,  {Duffy} C.~D.,   {Sutherland} J.~D.,  2015, \mn@doi [Nature Chemistry] {10.1038/nchem.2202}, \href {https://ui.adsabs.harvard.edu/abs/2015NatCh...7..301P} {7, 301}

\bibitem[\protect\citeauthoryear{Petrovic}{Petrovic}{2003}]{Petrovic2003}
Petrovic J.~J.,  2003, \mn@doi [Journal of Materials Science] {10.1023/A:1021134128038}, 38, 1

\bibitem[\protect\citeauthoryear{{Pierazzo} \& {Chyba}}{{Pierazzo} \& {Chyba}}{1999}]{PierazzoChyba1999}
{Pierazzo} E.,  {Chyba} C.~F.,  1999, \mn@doi [Meteoritics \& Planetary Science] {10.1111/j.1945-5100.1999.tb01409.x}, \href {https://ui.adsabs.harvard.edu/abs/1999M&PS...34..909P} {34, 909}

\bibitem[\protect\citeauthoryear{{Popova} et~al.,}{{Popova} et~al.}{2013}]{Popova2013}
{Popova} O.~P.,  et~al., 2013, \mn@doi [Science] {10.1126/science.1242642}, \href {https://ui.adsabs.harvard.edu/abs/2013Sci...342.1069P} {342, 1069}

\bibitem[\protect\citeauthoryear{{Popova}, {Borovi{\v{c}}ka}  \& {Campbell-Brown}}{{Popova} et~al.}{2019}]{Popova2019}
{Popova} O.,  {Borovi{\v{c}}ka} J.,   {Campbell-Brown} M.~D.,  2019, in {Ryabova} G.~O.,  {Asher} D.~J.,   {Campbell-Brown} M.~J.,  eds, , Meteoroids: Sources of Meteors on Earth and Beyond.
Cambridge University Press, p.~9

\bibitem[\protect\citeauthoryear{{Powner}, {Gerland}  \& {Sutherland}}{{Powner} et~al.}{2009}]{Powner2009}
{Powner} M.~W.,  {Gerland} B.,   {Sutherland} J.~D.,  2009, \mn@doi [\nat] {10.1038/nature08013}, \href {https://ui.adsabs.harvard.edu/abs/2009Natur.459..239P} {459, 239}

\bibitem[\protect\citeauthoryear{{Prialnik} \& {Jewitt}}{{Prialnik} \& {Jewitt}}{2022}]{PrialnikJewitt2022}
{Prialnik} D.,  {Jewitt} D.,  2022, \mn@doi [arXiv e-prints] {10.48550/arXiv.2209.05907}, \href {https://ui.adsabs.harvard.edu/abs/2022arXiv220905907P} {p. arXiv:2209.05907}

\bibitem[\protect\citeauthoryear{{Register}, {Mathias}  \& {Wheeler}}{{Register} et~al.}{2017}]{Register2017}
{Register} P.~J.,  {Mathias} D.~L.,   {Wheeler} L.~F.,  2017, \mn@doi [\icarus] {10.1016/j.icarus.2016.11.020}, \href {https://ui.adsabs.harvard.edu/abs/2017Icar..284..157R} {284, 157}

\bibitem[\protect\citeauthoryear{{Rimmer}}{{Rimmer}}{2023}]{Rimmer2023}
{Rimmer} P.~B.,  2023, \mn@doi [arXiv e-prints] {10.48550/arXiv.2305.04911}, \href {https://ui.adsabs.harvard.edu/abs/2023arXiv230504911R} {p. arXiv:2305.04911}

\bibitem[\protect\citeauthoryear{{Rimmer} \& {Shorttle}}{{Rimmer} \& {Shorttle}}{2024}]{RimmerShorttle2024}
{Rimmer} P.~B.,  {Shorttle} O.,  2024, \mn@doi [Life] {10.3390/life14040498}, \href {https://ui.adsabs.harvard.edu/abs/2024Life...14..498R} {14, 498}

\bibitem[\protect\citeauthoryear{{Sainsbury-Martinez} \& {Walsh}}{{Sainsbury-Martinez} \& {Walsh}}{2024}]{SainsburyMartinezWalsh2024}
{Sainsbury-Martinez} F.,  {Walsh} C.,  2024, \mn@doi [\apj] {10.3847/1538-4357/ad28b3}, \href {https://ui.adsabs.harvard.edu/abs/2024ApJ...966...39S} {966, 39}

\bibitem[\protect\citeauthoryear{{Sainsbury-Martinez}, {Walsh}  \& {Cooke}}{{Sainsbury-Martinez} et~al.}{2025}]{Sainsbury-Martinez2024}
{Sainsbury-Martinez} F.,  {Walsh} C.,   {Cooke} G.,  2025, \mn@doi [\apj] {10.3847/1538-4357/ad96ad}, \href {https://ui.adsabs.harvard.edu/abs/2025ApJ...982...29S} {982, 29}

\bibitem[\protect\citeauthoryear{{Sasselov}, {Grotzinger}  \& {Sutherland}}{{Sasselov} et~al.}{2020}]{Sasselov2020}
{Sasselov} D.~D.,  {Grotzinger} J.~P.,   {Sutherland} J.~D.,  2020, \mn@doi [Science Advances] {10.1126/sciadv.aax3419}, \href {https://ui.adsabs.harvard.edu/abs/2020SciA....6.3419S} {6, eaax3419}

\bibitem[\protect\citeauthoryear{{Sekanina}}{{Sekanina}}{1993}]{Sekanina1993}
{Sekanina} Z.,  1993, \mn@doi [Science] {10.1126/science.262.5132.382}, \href {https://ui.adsabs.harvard.edu/abs/1993Sci...262..382S} {262, 382}

\bibitem[\protect\citeauthoryear{Shoemaker}{Shoemaker}{1962}]{Shoemaker1962}
Shoemaker E.,  1962, in Kopal A.,  ed., , Physics and Astronomy of the Moon.
Academic Press, pp 283--571

\bibitem[\protect\citeauthoryear{{Silber}, {Boslough}, {Hocking}, {Gritsevich}  \& {Whitaker}}{{Silber} et~al.}{2018}]{Silber2018}
{Silber} E.~A.,  {Boslough} M.,  {Hocking} W.~K.,  {Gritsevich} M.,   {Whitaker} R.~W.,  2018, \mn@doi [Advances in Space Research] {10.1016/j.asr.2018.05.010}, \href {https://ui.adsabs.harvard.edu/abs/2018AdSpR..62..489S} {62, 489}

\bibitem[\protect\citeauthoryear{{Sinclair} \& {Wyatt}}{{Sinclair} \& {Wyatt}}{2022}]{Sinclair2022}
{Sinclair} C.~A.,  {Wyatt} M.~C.,  2022, \mn@doi [\mnras] {10.1093/mnras/stab3026}, \href {https://ui.adsabs.harvard.edu/abs/2022MNRAS.509..345S} {509, 345}

\bibitem[\protect\citeauthoryear{{Sinclair}, {Wyatt}, {Morbidelli}  \& {Nesvorn{\'y}}}{{Sinclair} et~al.}{2020}]{Sinclair2020}
{Sinclair} C.~A.,  {Wyatt} M.~C.,  {Morbidelli} A.,   {Nesvorn{\'y}} D.,  2020, \mn@doi [\mnras] {10.1093/mnras/staa3210}, \href {https://ui.adsabs.harvard.edu/abs/2020MNRAS.499.5334S} {499, 5334}

\bibitem[\protect\citeauthoryear{{Snodgrass}, {Fitzsimmons}, {Lowry}  \& {Weissman}}{{Snodgrass} et~al.}{2011}]{Snodgrass2011}
{Snodgrass} C.,  {Fitzsimmons} A.,  {Lowry} S.~C.,   {Weissman} P.,  2011, \mn@doi [\mnras] {10.1111/j.1365-2966.2011.18406.x}, \href {https://ui.adsabs.harvard.edu/abs/2011MNRAS.414..458S} {414, 458}

\bibitem[\protect\citeauthoryear{Sutherland}{Sutherland}{2016}]{Sutherland2016}
Sutherland J.~D.,  2016, \mn@doi [Angewandte Chemie International Edition] {https://doi.org/10.1002/anie.201506585}, 55, 104

\bibitem[\protect\citeauthoryear{{Svetsov}, {Nemtchinov}  \& {Teterev}}{{Svetsov} et~al.}{1995}]{Svetsov1995}
{Svetsov} V.~V.,  {Nemtchinov} I.~V.,   {Teterev} A.~V.,  1995, \mn@doi [\icarus] {10.1006/icar.1995.1116}, \href {https://ui.adsabs.harvard.edu/abs/1995Icar..116..131S} {116, 131}

\bibitem[\protect\citeauthoryear{{Todd} \& {{\"O}berg}}{{Todd} \& {{\"O}berg}}{2020}]{ToddOberg2020}
{Todd} Z.~R.,  {{\"O}berg} K.~I.,  2020, \mn@doi [Astrobiology] {10.1089/ast.2019.2187}, \href {https://ui.adsabs.harvard.edu/abs/2020AsBio..20.1109T} {20, 1109}

\bibitem[\protect\citeauthoryear{Todd, Wogan  \& Catling}{Todd et~al.}{2024}]{Todd2024}
Todd Z.~R.,  Wogan N.~F.,   Catling D.~C.,  2024, ACS Earth and Space Chemistry

\bibitem[\protect\citeauthoryear{{Tsai} et~al.,}{{Tsai} et~al.}{2023}]{Tsai2023}
{Tsai} S.-M.,  et~al., 2023, \mn@doi [\nat] {10.1038/s41586-023-05902-2}, \href {https://ui.adsabs.harvard.edu/abs/2023Natur.617..483T} {617, 483}

\bibitem[\protect\citeauthoryear{{Walton}, {Rigley}, {Lipp}, {Law}, {Suttle}, {Sch{\"o}nb{\"a}chler}, {Wyatt}  \& {Shorttle}}{{Walton} et~al.}{2024}]{Walton2024}
{Walton} C.~R.,  {Rigley} J.~K.,  {Lipp} A.,  {Law} R.,  {Suttle} M.~D.,  {Sch{\"o}nb{\"a}chler} M.,  {Wyatt} M.,   {Shorttle} O.,  2024, \mn@doi [Nature Astronomy] {10.1038/s41550-024-02212-z}, \href {https://ui.adsabs.harvard.edu/abs/2024NatAs...8..556W} {8, 556}

\bibitem[\protect\citeauthoryear{{Weissman}}{{Weissman}}{2007}]{Weissman2007}
{Weissman} P.~R.,  2007, in {Valsecchi} G.~B.,  {Vokrouhlick{\'y}} D.,   {Milani} A.,  eds,  IAU Symposium Vol. 236, Near Earth Objects, our Celestial Neighbors: Opportunity and Risk. pp 441--450, \mn@doi{10.1017/S1743921307003559}

\bibitem[\protect\citeauthoryear{{Wheeler}, {Register}  \& {Mathias}}{{Wheeler} et~al.}{2017}]{Wheeler2017}
{Wheeler} L.~F.,  {Register} P.~J.,   {Mathias} D.~L.,  2017, \mn@doi [\icarus] {10.1016/j.icarus.2017.02.011}, \href {https://ui.adsabs.harvard.edu/abs/2017Icar..295..149W} {295, 149}

\bibitem[\protect\citeauthoryear{{Wheeler}, {Mathias}, {Stokan}  \& {Brown}}{{Wheeler} et~al.}{2018}]{Wheeler2018}
{Wheeler} L.~F.,  {Mathias} D.~L.,  {Stokan} E.,   {Brown} P.~G.,  2018, \mn@doi [\icarus] {10.1016/j.icarus.2018.06.014}, \href {https://ui.adsabs.harvard.edu/abs/2018Icar..315...79W} {315, 79}

\bibitem[\protect\citeauthoryear{{Wogan}, {Catling}, {Zahnle}  \& {Lupu}}{{Wogan} et~al.}{2023}]{Wogan2023}
{Wogan} N.~F.,  {Catling} D.~C.,  {Zahnle} K.~J.,   {Lupu} R.,  2023, \mn@doi [The Planetary Science Journal] {10.3847/PSJ/aced83}, \href {https://ui.adsabs.harvard.edu/abs/2023PSJ.....4..169W} {4, 169}

\bibitem[\protect\citeauthoryear{{Young}, {Shahar}  \& {Schlichting}}{{Young} et~al.}{2023}]{Young2023}
{Young} E.~D.,  {Shahar} A.,   {Schlichting} H.~E.,  2023, \mn@doi [\nat] {10.1038/s41586-023-05823-0}, \href {https://ui.adsabs.harvard.edu/abs/2023Natur.616..306Y} {616, 306}

\bibitem[\protect\citeauthoryear{{Zahnle}}{{Zahnle}}{1992}]{Zahnle1992}
{Zahnle} K.~J.,  1992, \mn@doi [\jgr] {10.1029/92JE00787}, \href {https://ui.adsabs.harvard.edu/abs/1992JGR....9710243Z} {97, 10243}

\bibitem[\protect\citeauthoryear{{Zahnle} \& {Mac Low}}{{Zahnle} \& {Mac Low}}{1994}]{ZahnleMacLow1994}
{Zahnle} K.,  {Mac Low} M.-M.,  1994, \mn@doi [\icarus] {10.1006/icar.1994.1038}, \href {https://ui.adsabs.harvard.edu/abs/1994Icar..108....1Z} {108, 1}

\bibitem[\protect\citeauthoryear{{Zahnle}, {Pollack}, {Grinspoon}  \& {Dones}}{{Zahnle} et~al.}{1992}]{Zahnle1992_satellites}
{Zahnle} K.,  {Pollack} J.~B.,  {Grinspoon} D.,   {Dones} L.,  1992, \mn@doi [\icarus] {10.1016/0019-1035(92)90187-C}, \href {https://ui.adsabs.harvard.edu/abs/1992Icar...95....1Z} {95, 1}

\bibitem[\protect\citeauthoryear{{Zellner}, {McCaffrey}  \& {Butler}}{{Zellner} et~al.}{2020}]{Zellner2020}
{Zellner} N. E.~B.,  {McCaffrey} V.~P.,   {Butler} J. H.~E.,  2020, \mn@doi [Astrobiology] {10.1089/ast.2020.2216}, \href {https://ui.adsabs.harvard.edu/abs/2020AsBio..20.1377Z} {20, 1377}

\bibitem[\protect\citeauthoryear{{Zhang} et~al.,}{{Zhang} et~al.}{2023}]{Zhang2023}
{Zhang} Z.,  et~al., 2023, \mn@doi [\aj] {10.3847/1538-3881/acf768}, \href {https://ui.adsabs.harvard.edu/abs/2023AJ....166..198Z} {166, 198}

\bibitem[\protect\citeauthoryear{{de Niem}, {K{\"u}hrt}, {Morbidelli}  \& {Motschmann}}{{de Niem} et~al.}{2012}]{deNiem2012}
{de Niem} D.,  {K{\"u}hrt} E.,  {Morbidelli} A.,   {Motschmann} U.,  2012, \mn@doi [\icarus] {10.1016/j.icarus.2012.07.032}, \href {https://ui.adsabs.harvard.edu/abs/2012Icar..221..495D} {221, 495}

\makeatother
\end{thebibliography}


\appendix

\section{Model validation}
\label{appendix:chyba_validation}

\begin{figure}
    \centering
    \includegraphics[width=\linewidth]{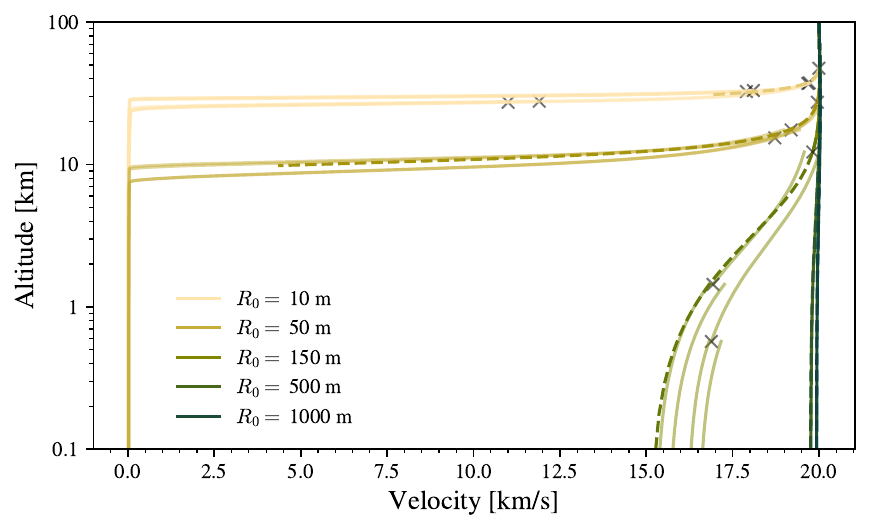}
    \caption{Here our numerical model (solid lines) is compared with the simple progressive fragmentation (`pancake') model described in \citet{Chyba1993} (dashed lines). The different colours correspond to initial comet size, as in figure~\ref{fig:comet_trajectory_gallery}. The integration is stopped when the comets radius increases by a factor of 6, at which point it is assumed the remaining mass and energy is deposited into the atmosphere in an airburst-type event \citep[see also;][]{Collins2005}.}
    \label{fig:chyba_comparison}
\end{figure}

First, we compare our results with the \citet{Chyba1993} pancake model; this is a widely-used pancake model that has been successfully validated against the 1908 Tunguska airburst. Figure~\ref{fig:chyba_comparison} demonstrates that our model is in good agreement, particularly at low ($\lesssim\,$100\,m) and large ($\gtrsim\,$500\,m) initial radii. Small deviations can be seen at intermediate sizes (or order 100\,m), which is driven by the onset of fragmentation, not included in the \citet{Chyba1993} model.

We can also, qualitatively, compare our results with the widely-used debris cloud model presented in \citet{HillsGoda1993}, in which all fragments continue their descent under a common bow shock. Their figure~4 records the height of half energy deposition as a function of initial meteor radius. For 10\,m comets, they find an altitude of approximately 40\,km, in good agreement with our figure~\ref{fig:comet_trajectory_gallery}. Comets larger than $1\,$km are found to retain more than 50\% of their initial kinetic energy, also in agreement with our model. Finally, their figure~6 records the fraction of mass ablated as a function of initial meteor radius. Comets smaller than 100\,m are found to lose all of their mass during atmospheric entry, which is also in good agreement with our model (see figures~\ref{fig:heatmaps} and~\ref{fig:minimum_cometary_diameter}).

\section{Comet trajectories in dense planetary atmospheres}
\label{appendix:comet_trajectories_dense_atmos}

\begin{figure*}
    \centering
    \includegraphics[width=\linewidth]{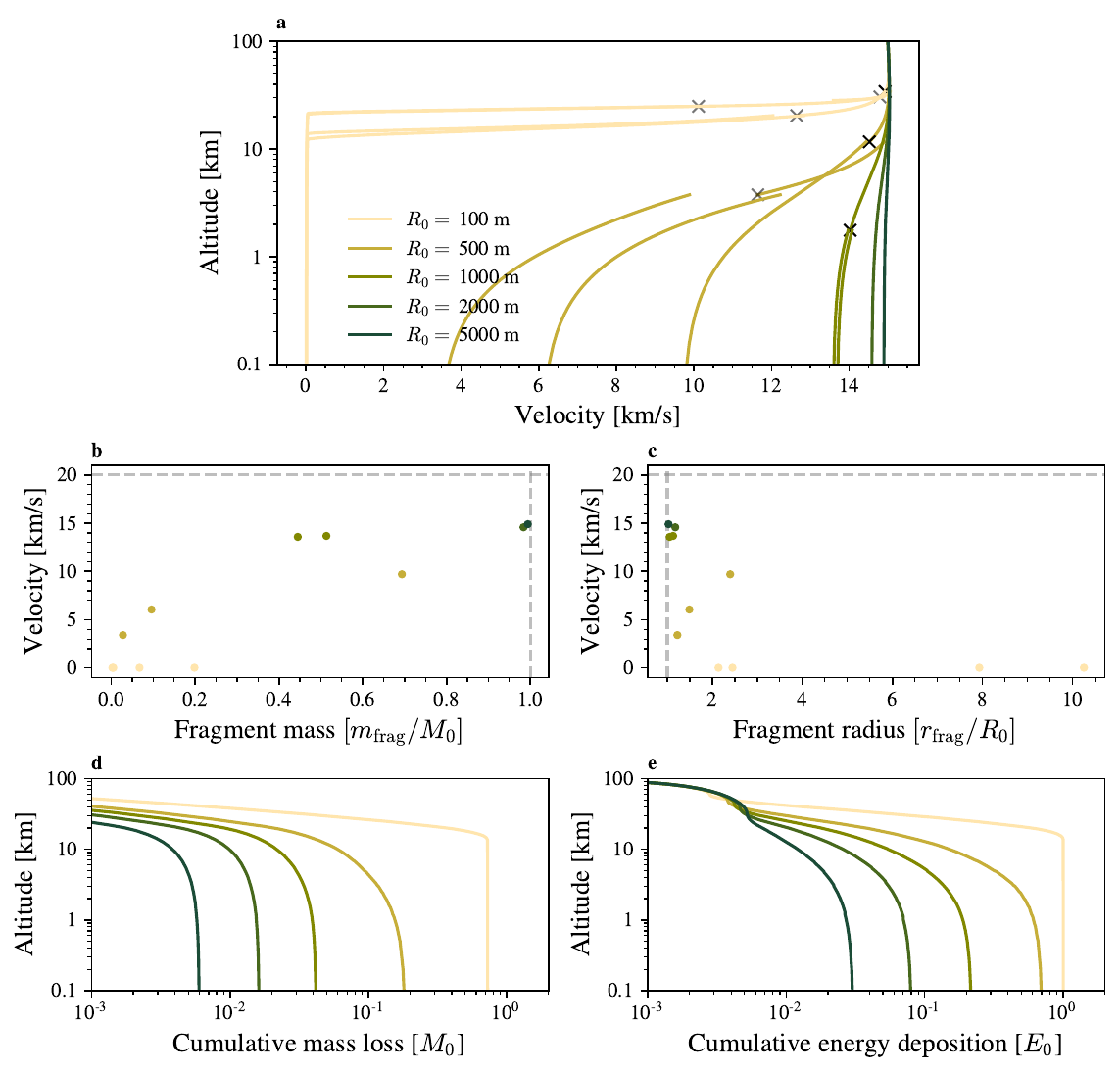}
    \caption{Same as figure~\ref{fig:comet_trajectory_gallery}, except for a dense planetary atmosphere, with surface density $12.25\,{\rm kg\,m}^{-3}$.}
    \label{fig:comet_trajectory_gallery_dense_atmos}
\end{figure*}

As discussed in \S\ref{sec:results_comet_fates} the interaction of small comets with dense atmospheres is particularly dramatic, given that the ram pressure experienced the leading edge of the comet is directly proportional to the atmospheric surface density. To first order, we expect analytically that the minimum cometary size able to survive atmospheric entry should by proportional to the atmospheric surface density (see equation~\ref{eq:qual_rcrit}). In figure~\ref{fig:comet_trajectory_gallery_dense_atmos} we show the atmospheric entry of comets into an atmosphere with density 10 times that of the modern Earth. As expected, much larger comets ($\sim\,$500\,m) are now significantly decelerated by the atmosphere, losing a much larger fraction of their initial mass to ablation. 

\newpage
\section{Simulation output data}
\label{appendix:tables}

There is variation in model results around $R_0=100\,$m, due to explicit stochasticity in the properties of child fragments, and the choice of $N_{\rm frag}$: the number of child fragments produced. In~\cref{tab:table_label2,tab:table_label3,tab:table_label4} we demonstrate the robustness of the numerical model to the choice of $N_{\rm frag}$, and list detailed simulation results for the most likely impact parameters (angle of incidence $45^\circ$ \citep{Shoemaker1962}, and impact velocity $\sim\,$20\,${\rm km\,s}^{-1}$ \citep{Weissman2007}). We report the following simulation outputs:
\begin{itemize}
    \item $R_0$ [km]: The initial radius of the comet at the top of the atmosphere.
    \item Fragmentation [True/False]: Whether the comet fragments in the atmosphere.
    \item $\langle m_{\rm imp}\rangle~[M_0]$: The average mass of fragments reaching the surface.
    \item $\langle v_{\rm imp}\rangle_m~[{\rm km\,s}^{-1}]$: The average (mass-weighted) impact velocity of fragments reaching the surface,
    \begin{equation}
        \langle v_{\rm imp}\rangle_m = \left(\sum_{i=1}^{N_{\rm frag, surf}}{m_iv_i}\right)\left(\sum_{i=1}^{N_{\rm frag, surf}}{m_i}\right)^{-1}.
    \end{equation}
    \item $N_{\rm frag, surf}$: The number of fragments that reach the surface.
\end{itemize}

Further output data are available directly from the \href{https://github.com/richard17a/atmosentry}{\textsc{atmosentry}} github repository, or by request.
\begin{table*}
    \centering
    \resizebox{0.7\textwidth}{!}{
    \begin{tabular}{|c||c||c||c||c|}
        \toprule
        $R_0$ [km] & Fragmentation & $\langle m_{\rm imp} \rangle~[M_0]$ & $\langle v_{\rm imp}~\rangle_m$ [${\rm km\,s}^{-1}$] & $N_{\rm frag, surf.}$ \\
        \midrule
        0.01 & True & 0.01 $\pm$ 0.00 & 0.01 $\pm$ 0.00 & 7.62 $\pm$ 0.48 \\
        0.02 & True & 0.01 $\pm$ 0.00 & 0.02 $\pm$ 0.00 & 7.62 $\pm$ 0.78 \\
        0.04 & True & 0.02 $\pm$ 0.00 & 0.02 $\pm$ 0.00 & 5.31 $\pm$ 0.85 \\
        0.05 & True & 0.02 $\pm$ 0.01 & 0.03 $\pm$ 0.00 & 5.00 $\pm$ 1.12 \\
        0.06 & True & 0.02 $\pm$ 0.01 & 0.03 $\pm$ 0.00 & 4.44 $\pm$ 0.86 \\
        0.08 & True & 0.03 $\pm$ 0.01 & 0.06 $\pm$ 0.06 & 4.19 $\pm$ 1.01 \\
        0.11 & True & 0.06 $\pm$ 0.03 & 6.67 $\pm$ 2.72 & 3.56 $\pm$ 0.61 \\
        0.14 & True & 0.18 $\pm$ 0.12 & 15.02 $\pm$ 0.51 & 3.56 $\pm$ 0.86 \\
        0.18 & True & 0.34 $\pm$ 0.06 & 17.88 $\pm$ 0.12 & 2.38 $\pm$ 0.48 \\
        0.23 & True & 0.45 $\pm$ 0.12 & 18.92 $\pm$ 0.02 & 2.06 $\pm$ 0.43 \\
        0.30 & True & 0.46 $\pm$ 0.00 & 19.36 $\pm$ 0.00 & 2.00 $\pm$ 0.00 \\
        0.39 & False & 0.95 $\pm$ 0.00 & 19.61 $\pm$ 0.00 & 1.00 $\pm$ 0.00 \\
        0.50 & False & 0.97 $\pm$ 0.00 & 19.76 $\pm$ 0.00 & 1.00 $\pm$ 0.00 \\
        0.60 & False & 0.97 $\pm$ 0.00 & 19.83 $\pm$ 0.00 & 1.00 $\pm$ 0.00 \\
        0.96 & False & 0.98 $\pm$ 0.00 & 19.92 $\pm$ 0.00 & 1.00 $\pm$ 0.00 \\
        1.53 & False & 0.99 $\pm$ 0.00 & 19.97 $\pm$ 0.00 & 1.00 $\pm$ 0.00 \\
        2.45 & False & 0.99 $\pm$ 0.00 & 20.00 $\pm$ 0.00 & 1.00 $\pm$ 0.00 \\
        3.91 & False & 1.00 $\pm$ 0.00 & 20.02 $\pm$ 0.00 & 1.00 $\pm$ 0.00 \\
        6.26 & False & 1.00 $\pm$ 0.00 & 20.03 $\pm$ 0.00 & 1.00 $\pm$ 0.00 \\
        10.00 & False & 1.00 $\pm$ 0.00 & 20.04 $\pm$ 0.00 & 1.00 $\pm$ 0.00 \\
        \bottomrule
    \end{tabular}}
    \caption{Number of child fragments per fragment event, $N_{\rm frag} = 2$. Initial velocity $v_0 = 20\,{\rm km\,s}^{-1}$.}
    \label{tab:table_label2}
\end{table*}

\begin{table*}
    \centering
    \resizebox{0.7\textwidth}{!}{
    \begin{tabular}{|c||c||c||c||c|}
        \toprule
        $R_0$ [km] & Fragmentation & $\langle m_{\rm imp} \rangle~[M_0]$ & $\langle v_{\rm imp}~\rangle_m$ [${\rm km\,s}^{-1}$] & $N_{\rm frag, surf.}$ \\
        \midrule
        0.01 & True & 0.01 $\pm$ 0.00 & 0.01 $\pm$ 0.00 & 7.56 $\pm$ 0.70 \\
        0.02 & True & 0.01 $\pm$ 0.00 & 0.02 $\pm$ 0.00 & 7.50 $\pm$ 0.87 \\
        0.04 & True & 0.02 $\pm$ 0.01 & 0.02 $\pm$ 0.00 & 5.50 $\pm$ 1.17 \\
        0.05 & True & 0.02 $\pm$ 0.00 & 0.03 $\pm$ 0.00 & 5.25 $\pm$ 0.83 \\
        0.06 & True & 0.02 $\pm$ 0.01 & 0.03 $\pm$ 0.00 & 4.56 $\pm$ 1.12 \\
        0.08 & True & 0.02 $\pm$ 0.00 & 0.04 $\pm$ 0.02 & 4.19 $\pm$ 0.63 \\
        0.11 & True & 0.05 $\pm$ 0.02 & 6.16 $\pm$ 2.44 & 3.75 $\pm$ 0.56 \\
        0.14 & True & 0.19 $\pm$ 0.12 & 15.14 $\pm$ 0.52 & 3.44 $\pm$ 0.79 \\
        0.18 & True & 0.40 $\pm$ 0.17 & 17.86 $\pm$ 0.15 & 2.25 $\pm$ 0.66 \\
        0.23 & True & 0.44 $\pm$ 0.00 & 18.92 $\pm$ 0.00 & 2.00 $\pm$ 0.00 \\
        0.30 & True & 0.46 $\pm$ 0.00 & 19.36 $\pm$ 0.00 & 2.00 $\pm$ 0.00 \\
        0.39 & False & 0.95 $\pm$ 0.00 & 19.61 $\pm$ 0.00 & 1.00 $\pm$ 0.00 \\
        0.50 & False & 0.97 $\pm$ 0.00 & 19.76 $\pm$ 0.00 & 1.00 $\pm$ 0.00 \\
        0.60 & False & 0.97 $\pm$ 0.00 & 19.83 $\pm$ 0.00 & 1.00 $\pm$ 0.00 \\
        0.96 & False & 0.98 $\pm$ 0.00 & 19.92 $\pm$ 0.00 & 1.00 $\pm$ 0.00 \\
        1.53 & False & 0.99 $\pm$ 0.00 & 19.97 $\pm$ 0.00 & 1.00 $\pm$ 0.00 \\
        2.45 & False & 0.99 $\pm$ 0.00 & 20.00 $\pm$ 0.00 & 1.00 $\pm$ 0.00 \\
        3.91 & False & 1.00 $\pm$ 0.00 & 20.02 $\pm$ 0.00 & 1.00 $\pm$ 0.00 \\
        6.26 & False & 1.00 $\pm$ 0.00 & 20.03 $\pm$ 0.00 & 1.00 $\pm$ 0.00 \\
        10.00 & False & 1.00 $\pm$ 0.00 & 20.04 $\pm$ 0.00 & 1.00 $\pm$ 0.00 \\
        \bottomrule
    \end{tabular}}
    \caption{Number of child fragments per fragment event, $N_{\rm frag} = 3$. Initial velocity $v_0 = 20\,{\rm km\,s}^{-1}$.}
    \label{tab:table_label3}
\end{table*}

\begin{table*}
    \centering
    \resizebox{0.7\textwidth}{!}{
    \begin{tabular}{|c||c||c||c||c|}
        \toprule
        $R_0$ [km] & Fragmentation & $\langle m_{\rm imp} \rangle~[M_0]$ & $\langle v_{\rm imp}~\rangle_m$ [${\rm km\,s}^{-1}$] & $N_{\rm frag, surf.}$ \\
        \midrule
        0.01 & True & 0.01 $\pm$ 0.00 & 0.01 $\pm$ 0.00 & 7.88 $\pm$ 0.33 \\
        0.02 & True & 0.01 $\pm$ 0.00 & 0.02 $\pm$ 0.00 & 7.38 $\pm$ 1.05 \\
        0.04 & True & 0.02 $\pm$ 0.00 & 0.02 $\pm$ 0.00 & 5.62 $\pm$ 0.78 \\
        0.05 & True & 0.02 $\pm$ 0.00 & 0.03 $\pm$ 0.00 & 5.00 $\pm$ 0.71 \\
        0.06 & True & 0.02 $\pm$ 0.01 & 0.03 $\pm$ 0.01 & 4.38 $\pm$ 0.93 \\
        0.08 & True & 0.02 $\pm$ 0.00 & 0.05 $\pm$ 0.04 & 4.25 $\pm$ 0.75 \\
        0.11 & True & 0.06 $\pm$ 0.03 & 6.57 $\pm$ 2.20 & 3.44 $\pm$ 0.61 \\
        0.14 & True & 0.14 $\pm$ 0.04 & 14.88 $\pm$ 0.35 & 3.94 $\pm$ 0.56 \\
        0.18 & True & 0.39 $\pm$ 0.12 & 17.85 $\pm$ 0.11 & 2.19 $\pm$ 0.53 \\
        0.23 & True & 0.42 $\pm$ 0.05 & 18.92 $\pm$ 0.01 & 2.12 $\pm$ 0.33 \\
        0.30 & True & 0.46 $\pm$ 0.00 & 19.36 $\pm$ 0.00 & 2.00 $\pm$ 0.00 \\
        0.39 & False & 0.95 $\pm$ 0.00 & 19.61 $\pm$ 0.00 & 1.00 $\pm$ 0.00 \\
        0.50 & False & 0.97 $\pm$ 0.00 & 19.76 $\pm$ 0.00 & 1.00 $\pm$ 0.00 \\
        0.60 & False & 0.97 $\pm$ 0.00 & 19.83 $\pm$ 0.00 & 1.00 $\pm$ 0.00 \\
        0.96 & False & 0.98 $\pm$ 0.00 & 19.92 $\pm$ 0.00 & 1.00 $\pm$ 0.00 \\
        1.53 & False & 0.99 $\pm$ 0.00 & 19.97 $\pm$ 0.00 & 1.00 $\pm$ 0.00 \\
        2.45 & False & 0.99 $\pm$ 0.00 & 20.00 $\pm$ 0.00 & 1.00 $\pm$ 0.00 \\
        3.91 & False & 1.00 $\pm$ 0.00 & 20.02 $\pm$ 0.00 & 1.00 $\pm$ 0.00 \\
        6.26 & False & 1.00 $\pm$ 0.00 & 20.03 $\pm$ 0.00 & 1.00 $\pm$ 0.00 \\
        10.00 & False & 1.00 $\pm$ 0.00 & 20.04 $\pm$ 0.00 & 1.00 $\pm$ 0.00 \\
        \bottomrule
    \end{tabular}}
    \caption{Number of child fragments per fragment event, $N_{\rm frag} = 4$. Initial velocity $v_0 = 20\,{\rm km\,s}^{-1}$.}
    \label{tab:table_label4}
\end{table*}


\bsp	
\label{lastpage}
\end{document}